\documentclass[times,twocolumn,final]{elsarticle}

\usepackage{graphicx}
\graphicspath{ {blend/} }
\usepackage{algorithm}
\usepackage[noend]{algpseudocode}
\usepackage[format=plain,
labelfont={bf,it},
textfont=it]{caption}
\usepackage{subfig}
\usepackage{wrapfig}
\usepackage{booktabs}

\usepackage[utf8]{inputenc}
\usepackage[T1]{fontenc}
\usepackage{amsmath}

\usepackage{cag}
\usepackage{framed,multirow}

\usepackage{amssymb}
\usepackage{latexsym}

\usepackage{url}
\usepackage{xcolor}
\definecolor{newcolor}{rgb}{.8,.349,.1}

\usepackage{hyperref}

\usepackage[switch,pagewise]{lineno} 

\newcommand{\todo}[1]{}
\renewcommand{\todo}[1]{{\color{red} TODO: {#1}}}

\journal{Computers \& Graphics}

\begin{document}

\verso{Preprint Submitted for review}

\begin{frontmatter}
\title{Deep Radiance Caching: Convolutional Autoencoders Deeper in Ray Tracing}
\author[1]{Giulio \snm{Jiang}}

\emailauthor{b.kainz@imperial.ac.uk}{Bernhard Kainz}
\ead{giuliojiang@gmail.com}

\author[1]{Bernhard \snm{Kainz}\corref{cor1}}
\cortext[cor1]{Corresponding author: }

\address[1]{Imperial College London, SW7 2AZ, London}
\received{\today}

\begin{abstract}
Rendering realistic images with global illumination is a computationally demanding
task and often requires dedicated hardware for feasible runtime. Recent research uses 
Deep Neural Networks to predict indirect lighting on image level,
but such methods are commonly limited to diffuse materials and require training
on each scene.
We present Deep Radiance Caching (DRC), an efficient variant of Radiance Caching utilizing Convolutional Autoencoders for rendering global illumination. 
DRC employs a denoising neural network with Radiance Caching to
support a wide range of material types,
without the requirement of offline pre-computation or training for each scene. This offers high performance CPU rendering for maximum accessibility.
Our method has been evaluated on interior scenes, and is able to produce high-quality images
within 180 seconds on a single CPU.
\end{abstract}


\end{frontmatter}


\section{Introduction}

 Ray Tracing is capable of producing photo-realistic images virtually indistinguishable
from real pictures. Progressive refinements on rendering algorithms, such as
Bi-Directional Path Tracing (BDPT) \cite{lafortune1993bi} and Metropolis Light Transport
\cite{veach1997metropolis} have increased the efficiency of rendering engines in scenarios
in which light paths are difficult to evaluate due to the high amount of indirect lighting and Global Illumination.
Complex lighting conditions are, however, still highly expensive to resolve, and most
algorithms require long rendering times to reduce the noise from Monte Carlo sampling.

\begin{figure}[!htbp]
	\vspace{-0.45cm}
	\includegraphics[width=\linewidth]{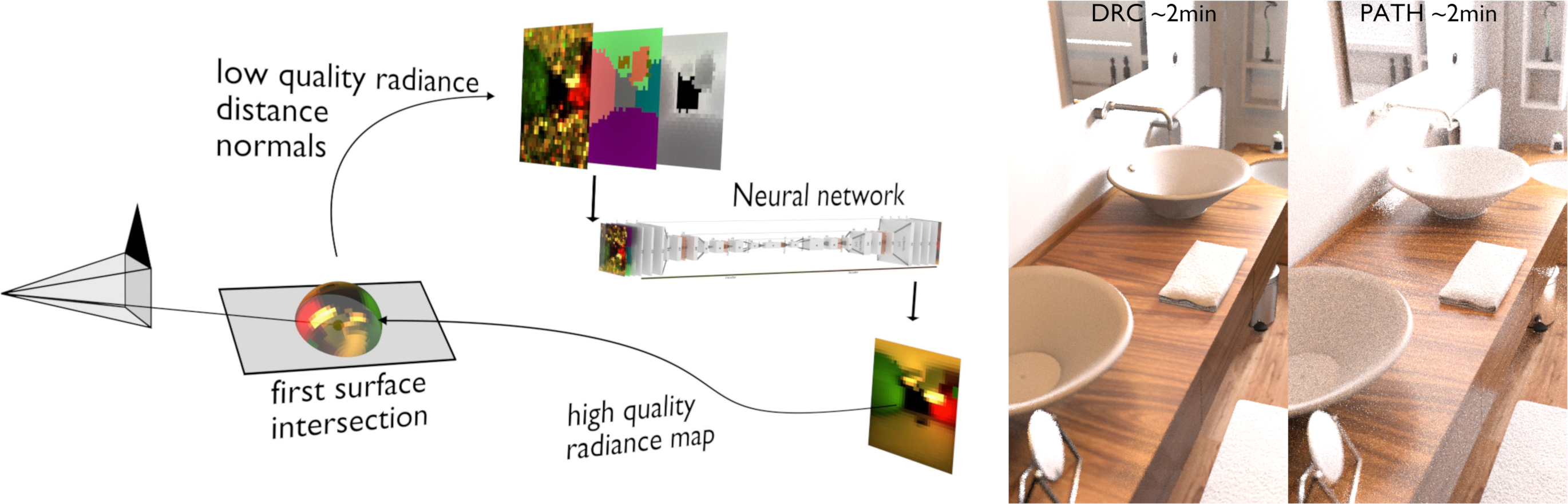}
	\centering
	\caption{
		Our proposed Deep Radiance Caching (DRC) method evaluates a low quality radiance map, distance and
		normals at the first intersection, and uses a neural network to obtain a higher
		quality radiance map containing all indirect lighting. The \textit{Bathroom} scene
		from PBRTv3's scenes~\cite{pbrt_v3_scenes} shows DRC compared
		to a same-time path traced image. DRC produces a noise-free result with convincing
		global illumination after 30 seconds, and progressively refines quality afterwards.
	}
	\label{fig:teaser}
	\vspace{-0.3cm}
\end{figure}

 Biased methods have been implemented to produce convincing quality images at a fraction
of the cost required by a ray tracer. An early method, Instant Radiosity \cite{keller1997instant},
exploited the low rate of illumination change over diffuse surfaces to approximate global illumination
by rendering the same scene many times using Virtual Point Lights sampled at locations
reached by the main light sources to simulate secondary bounces.

Radiance Caching \cite{krivanek2005radiance} focuses on accelerating rendering of
glossy materials, by caching an optimized representation of the radiance received
on a surface. This method enables view-dependant reflections to be rendered correctly.

Biased algorithms have evolved and recent research has attempted
to use machine learning techniques to accelerate rendering
of global illumination effects. The \textit{Deep Illumination} \cite{thomas_deep_illumination} approach
uses a GAN to translate diffuse albedos, normals and depth maps to a global illumination
component layer and obtained good results at predicting indirect illumination in real
time for diffuse materials. The network requires specific training for each scene to be
rendered, but is able to extrapolate and adapt to dynamic objects and newly introduced shapes.

To overcome the limitations of diffuse-only materials of current machine learning
renderers, and to make the renderer easily
usable without the need for offline training, we attempt a more general approach
to resolving global illumination. Based on the estimation and caching of radiance maps for indirect illumination only, we combine Radiance Caching and a neural network that
reconstructs and denoises radiance maps.
Indirect lighting is the primary cause of strong noise in Monte Carlo rendering. This noise is much more difficult to clear than first bounce lighting due to the high dimensionality
of the paths. Rendering can therefore be accelerated by predicting approximate
but noise-free radiance maps that can be interpolated and used to obtain the indirect
illumination component of the final image. Thanks to the slowly changing nature of indirect
illumination, artifacts and bias are not very visible in the general case, while the
overall predicted global illumination looks convincing and noise-free.

In contrast to full image level methods, Deep Radiance Caching (DRC) as outlined in Figure~\ref{fig:teaser} uses path tracing to find the first intersection in the scene. A
Convolutional Autoencoder predicts at this intersection a high quality, geometry-dependent radiance map from a path traced
map rendered at just 1 sample per pixel, a 
depth and a normal map. 

Using high quality radiance maps we
can approximate all indirect illumination contributed from path tracing bounces beyond the
first one.
We use a neural network to predict high quality
radiance maps for these approximations. Consequently, we accelerate the rendering of global illumination effects significantly. Our approach works with a wide range
of material types, and does not require any offline pre-computation or per-scene
training.

\section{Related Work} \label{sec:relatedWork}

\subsection{Advanced denoising systems} 

Recent research uses image denoising and machine learning techniques for rendering and image generation purposes. 
The most immediate way to apply machine learning to graphics and ray tracing is to
operate on the final rendered image level. These approaches take as input a scene
rendered from the final camera's viewpoint, and attempt to output a transformation
that results in higher visual quality, removal of noise, or the addition of effects. 

Denoising of natural images is a core research area in Computer Vision. Numerous approaches have been proposed using, \emph{e.g.}, Total Variation~\cite{chambolle2011first,beck2009fast}, Non-local means filtering~\cite{buades2005non}, dictionary learning~\cite{elad2006image} or recently, wavelet transformations of the contracting path of a convolutional neural auto-encoder with skip connections to gain a higher perceptive field with minimal computational costs~\cite{liu2018multi}. 

Monte Carlo rendering methods like path tracing produce images with stochastic noise artifacts, thus image denoising algorithms are a natural fit for improving the final image quality.   
Denoising of Monte Carlo renderings has been intensely studied in Computer Graphics literate. For example, \cite{bitterli2016nonlinearly} use a first-order model with auxiliary buffers and a Non-local means nonlinear regression kernel to predict optimal filter parameters. \cite{moon2016adaptive} proposes an adaptive rendering method, which fits local polynomial functions to approximate the image and predicts the local optimal polynomial order with a multistage error estimation process. 

One of the first neural network-based approaches to improve rendering quality
\cite{kalantari_machine_learning_ml_noise} used a network to obtain
filter parameters to denoise path traced images. The method relies on collecting
primary features such as world coordinates, surface normals, texture values and
illumination visibility from a ray tracer, from which secondary features are
computed: gradients, mean and deviation. A Multi Layer Perceptron (MLP)
uses the secondary features to output filter parameters. The method achieves good denoising
quality, but requires the use of a modified ray tracer-based on PBRTv2 \cite{pbrt2}.
\textit{Kernel-predicting convolutional networks for denoising Monte Carlo renderings} 
\cite{kernel_predicting_denoising_2017} and the recent extension
\textit{Denoising with Kernel Prediction and Asymmetric Loss Functions} \cite{Vogels2018KPAL}
improve denoising networks
by separating the processing of diffuse and specular layers
with Convolutional Neural Networks (CNN).
This approach is also highly applicable to existing ray tracers, as many can natively
output separate layers for diffuse, specular, z-buffer, mist and others. The separation
of layers allows the network to better handle the high dynamic range of input values. The introduction of logarithmic transforms to the input data further increases
performance. The CNN shows excellent performance with moderate noise.

A method proposed by Chaitanya et al 
\cite{chaitanya2017_interactive_reconstruction} differs from
the other examples as it does not take inputs with a small number of samples per pixel,
but with exactly one sample. By retaining information from previous frames in an animation,
the used Convolutional Autoencoder is capable of reconstructing missing details given a very
small amount of information. This method is particularly suited for fast,
interactive animation previews.
A variant of this approach has been proposed by~\cite{kuznetsov2018deep} to tackle the problem of image reconstruction from sparsely, non-uniformly sampled renderings. 

Very recent approaches focus on different aspects of image-level denoising, like multi-variate pixel values with color information from different scene depths~\cite{vicini2019denoising}, novel denoising architecture to relate permutation variant samples directly to the output image through splatting~\cite{gharbi2019sample}, and \cite{xu2019adversarial} who use the power of generative adversarial networks to introduce  more realistic high-frequency details and global illumination by learning the distribution from a set of high-quality Monte Carlo path tracing images. The latter is related to popular super-resolution techniques in the Computer Vision community, \emph{e.g.}~\cite{ledig2017photo}. 


%

The above approaches are advanced denoising systems, which are limited regarding the details they can create that were not captured by the rendering engine or missing due
to the low quality of the output. Missing details that were not captured in the current frame can only be reconstructed from information present in previously rendered frames, \emph{e.g.}, \cite{chaitanya2017_interactive_reconstruction}. 


\subsection{Global illumination with machine learning} In the following, we highlight some past research that focuses on adding Indirect Illumination and
global illumination effects using neural networks by operating on the image plane level.

A CNN implementation in \textit{Deep Shading} \cite{nalbach_deep_shading} 
predicts screen-space effects such as Ambient Occlusion, Motion Blur, Anti-Aliasing 
and Diffuse Indirect Illumination.
The CNN takes OpenGL rasterization primary features, and produces
outputs that can match real-time algorithms typically used in video games, while being
able to generate any combination of effects in a single pass. This approach does
not attempt to achieve photorealistic results. 

A Conditional Generative Adversarial Network
(CGAN)  is trained in \textit{Deep Illumination} \cite{thomas_deep_illumination} 
on specific scenes, with diffuse surfaces, to generate realistic and accurate global illumination 
effects. Despite being trained on specific scenes, the network is able to deal with
moving objects and newly introduced shapes. Its temporal stability makes it suitable
for producing animations, given a limited scene variability.

\textit{Global illumination with radiance regression functions} \cite{ren2013global} focuses on
realtime indirect illumination rendering. This approach uses a neural network that learns the relationship
between local and contextual attributes such as vertices and light position, to the indirect
illumination value. This method shows very good performance and quality, although it is
limited to point light sources and requires pre-baking of the Radiance Regression Function
for each scene.

\subsection{Machine learning for integration} 
A different integration point for machine learning is to accelerate the convergence of existing path tracers and bidirectional path tracers.
Reinforcement Learning (RL) \cite{dahm2016learning} exploits the similarity between the
rendering equation and RL.
The 3D scene is subdivided into Voronoi cells, each with an importance map dome that is updated
using RL to remember the most efficient light path directions.
The system is a path tracer that learns to dynamically update local importance sampling
maps to increase its path efficiency and to reduce the number of zero-contribution paths. A similar idea is presented in \textit{Machine Learning and Integral Equations}~\cite{integralEquations}, where an approximate solution to the integral equation is learned.

\textit{On-line Learning of Parametric Mixture Models for Light Transport Simulation} \cite{parametricMixtureLt} optimizes the learning of difficult sampling distributions to guide
standard ray tracing algorithms. This significantly improves the importance accuracy of light rays.

\textit{Practical Path  Guiding}~\cite{herholz2016product,muller2017practical,mueller19guiding} improves the efficiency of path tracing by
adjusting importance sampling of direct and indirect bounces based on the radiance received on
surfaces of the scene. This method accelerates initial convergence of ray tracing in difficult lighting
situations and has been adopted in production engines.

\textit{Deep Scattering} \cite{kallweit2017deep} shows the power of neural networks when applied to a very specific use case:
rendering of clouds. This method is able to achieve extremely high quality in the generation of complex
lighting effects such as cloud silverlining by using a hierarchical 3D representation of the cloud structure.

\section{Contribution}

We propose Deep Radiance Caching (DRC), an efficient variant of radiant caching, which separates the processing of direct and indirect lighting. DRC is able to produce convincing indirect
illumination, including glossy reflections and ambient occlusion, without any
preprocessing on the scene. Our main contributions are:

\begin{itemize}
	
	\item Use of a Convolutional Autoencoder to predict high quality radiance maps
	from low quality data for indirect illumination;
	
	\item Reintegration of direct illumination using a separate integrator;
	
	\item Progressive sampling and smooth interpolation of indirect illumination values
	without pre-computation and additional storage.
	
\end{itemize}

\section{Method}


DRC renders the final image by computing direct and indirect illumination independently.
The direct illumination pass uses a standard ray tracer with depth fixed at 1.
The indirect illumination subdivides the image into tiles, and computes several radiance maps within each tile. Each radiance map is obtained by ray tracing a low resolution intensity, depth and normals map from the first intersection point into the scene. These maps are augmented by the Convolutional Autoencoder. The radiance maps are used to compute the final indirect illumination contribution of the pixels.


\subsection{Illumination components}


DRC splits rendering of direct illumination and indirect illumination.
The indirect illumination component evaluates cached radiance maps, obtained efficiently using
a deep neural network that reads local geometrical information of the scene.
A standard path tracer evaluates the Rendering Equation \cite{kajiya1986rendering}
recursively:
\begin{equation}\label{eq:renderingEquation}
\begin{aligned}
& L_{o}(x, \vec{\omega}_{0}) = L_{e}(x, \vec{\omega}_{0}) + \\ +
& \int_{\Omega _{2\pi }}^{{}} f_{r}(x, \vec{\omega}_{0}, \vec{\omega}_{i}) cos(\theta_{i}) L_{i}(x, \vec{\omega}_{i}) d \omega_{i}.
\end{aligned}
\end{equation}

$L_{e}$ is the radiance emitted by the surface, $f_{r}$ is the BRDF or BSDR of the material,
the $cos$ term is used to compute the irradiance accounting for the incident angle of the incoming radiance $L_{i}$ from
the next bounce. A ray in path tracing
bounces on surfaces in the scene until a light is hit, or the path is terminated.
The recursively evaluated radiance, the BRDF $f$ of the material
and the viewing direction $\vec{\omega}_{i}$ are used to obtain the final pixel value.
DRC approximates the evaluation of the Rendering
Equation by collapsing all the light bounces beyond the first into a single step.
A ray tracer shoots a ray to find the first intersection point, and evaluates
on its hemisphere a predicted radiance map, a depth map, and a normal map. These three
images are the inputs for our Convolutional Autoencoder, which returns a higher
quality radiance map via

\begin{equation}\label{eq:renderingEquation2}
\begin{aligned}
L_{o}(x, \vec{\omega}_{0}) &= L_{e}(x, \vec{\omega}_{0}) + \\
&\int_{\Omega _{2\pi }}^{{}} f_{r}(x, \vec{\omega}_{0}, \vec{\omega}_{i}) cos(\theta_{i}) R_{i}(\vec{\omega}_{i}) d \omega_{i}.
\end{aligned}
\end{equation}

We replace the recursive radiance term with $R$, the high quality radiance map
obtained from the neural network, and evaluate it in the same way according
to the BRDF $f$ of the surface in the viewing direction $\vec{\omega}_{i}$.

Previous research has achieved excellent degrees of success in denoising
and reconstructing missing parts of the image from the output of a path
tracer using Convolutional Neural Networks. Our network is
particularly suited for extracting high level information from the locality
of the image data. In the context of DRC, we chose to implement a
Convolutional Autoencoder, as the noisy input can be associated to a highly
lossy and compressed version of the same image when rendered at a higher
number of samples per pixel.
The output radiance map is placed back on the hemisphere
around the first intersection point, and contains an approximation of all the
recursive indirect radiance that path tracing would evaluate over time. 

In practice, $R$ in Equation \ref{eq:renderingEquation2} does not have infinite resolution. 
DRC is able to preserve the high level of detail of the first-bounce illumination by
excluding the direct light source contribution from $R$, and reintroducing it
using a separate direct illumination pass. The final DRC Rendering Equation is

\begin{equation}\label{eq:renderingEquation3}
\begin{aligned}
L_{o}(x, \vec{\omega}_{0}) &= L_{e}(x, \vec{\omega}_{0}) + \\
&\int_{\Omega _{2\pi }}^{{}} f_{r}(x, \vec{\omega}_{0}, \vec{\omega}_{i}) cos(\theta_{i}) R_{i}(\vec{\omega}_{i}) d \omega_{i} + \\
&\int_{\Omega _{2\pi }}^{{}} f_{r}(x, \vec{\omega}_{0}, \vec{\omega}_{i}) cos(\theta_{i}) D_{i}(\vec{\omega}_{i}) d \omega_{i}.
\end{aligned}
\end{equation}

$D_{i}$ is the direct illumination contribution from light sources visible at the first bounce.
DRC can use a low resolution radiance map to approximate the $R$ term, while the direct illumination rendering pass
preserves high frequency details such as edges and hard shadows.

Evaluating the entire radiance map $R$ would still be expensive. Our implementation
uses Multiple Importance Sampling \cite{veach1995bidirectional_estimators} with a small
fixed number of samples to obtain indirect illumination values. We alternate sampling
from $R$ as if it were an environment map with perfect
visibility, and from the probability distribution of the BRDF.
Ambient occlusion is encoded into these local environment maps because they are taken from
the point of intersection.

To include direct illumination, the first intersection point also generates shadow
rays to light sources (not shown in Figure \ref{fig:teaser}). It is important to exclude
direct light information from the radiance maps, as they do not have a high enough resolution
 to produce accurate shadows.

\subsection{Progressive refinement and interpolation} \label{sec:progressRefinInter}

Instead of evaluating a radiance map at each pixel of the image, DRC accelerates
the rendering process by sampling a few pixels, interpolating their indirect
illumination values, and progressively refining the final picture by adding
samples. Sampling and interpolation only applies to indirect illumination, the
direct illumination passes are performed over all pixels.

Progressive refinement allows the user to start rendering without setting a target number of passes,
but to watch the image as its quality improves and stop the process once satisfied.
Additional progressive refinement passes are also effective for  smoothing small inconsistencies in indirect
illumination.

The interpolation method of DRC does not use any precomputation to determine
the best sampling locations. We compute a grid of sampling points on the image plane,
evaluate a radiance map at each point, and interpolate all other values. 
After each complete pass on the image, the distance between points on the grid 
is reduced by a constant factor, allowing the algorithm to progressively refine 
the resolution of the indirect illumination component.

The interpolation strategy for points that are not sampled is based on both the
distance from the sampled radiance
maps and the surface normal at the primary intersection.
The primary intersection is the closest non-specular intersection point of a ray
shot from the camera. If a ray hits a transmission or mirror material, we follow
the bounce recursively, and place the primary intersection point when the bounced
ray hits again.
The weight of each radiance map for pixel $i$ is based on the  simple heuristic
\begin{equation}
w = w_p \cdot w_n + w_p + \epsilon.
\end{equation}

Position and normal weights are multiplied to increase the weight when both are high,
and position weight is added again to generate smoother weight values when the normal weight changes
rapidly in noisy areas.
$\epsilon$ is a small term used to avoid computation errors when all other terms are zero. 
The position weight 
is  
\begin{equation}
w_p = dist(p_i, p) / r.
\end{equation} 
$dist(p_i, p)$ measures the distance in pixels on the image plane between
the pixel being evaluated and the cached radiance map, and
$r$ is the distance between two points in the grid.

The normal weight $w_n$ is maximal when the pixel's intersection point
and the cached radiance map's normal (~$n$~ and ~$n_i$~ respectively) point 
in the same direction, and becomes zero when their dot product is negative:
\begin{equation}
w_n = 1 -
max \left( 0,
\left( \frac{n_i}{\left| n_i \right|} \cdot \frac{n}{\left| n \right|} \right)
\right)
\end{equation}

The weights $w$ of the radiance maps are normalized and converted into
sampling probabilities for the radiance sampling.

\subsection{Neural network}

A neural network is at the core of DRC. The network predicts a smooth and accurate radiance map from primary features
that can be computed quickly by a first hit ray tracer.

\begin{figure}[!htbp]
	\begin{center}
		\includegraphics[width=1.0\columnwidth]{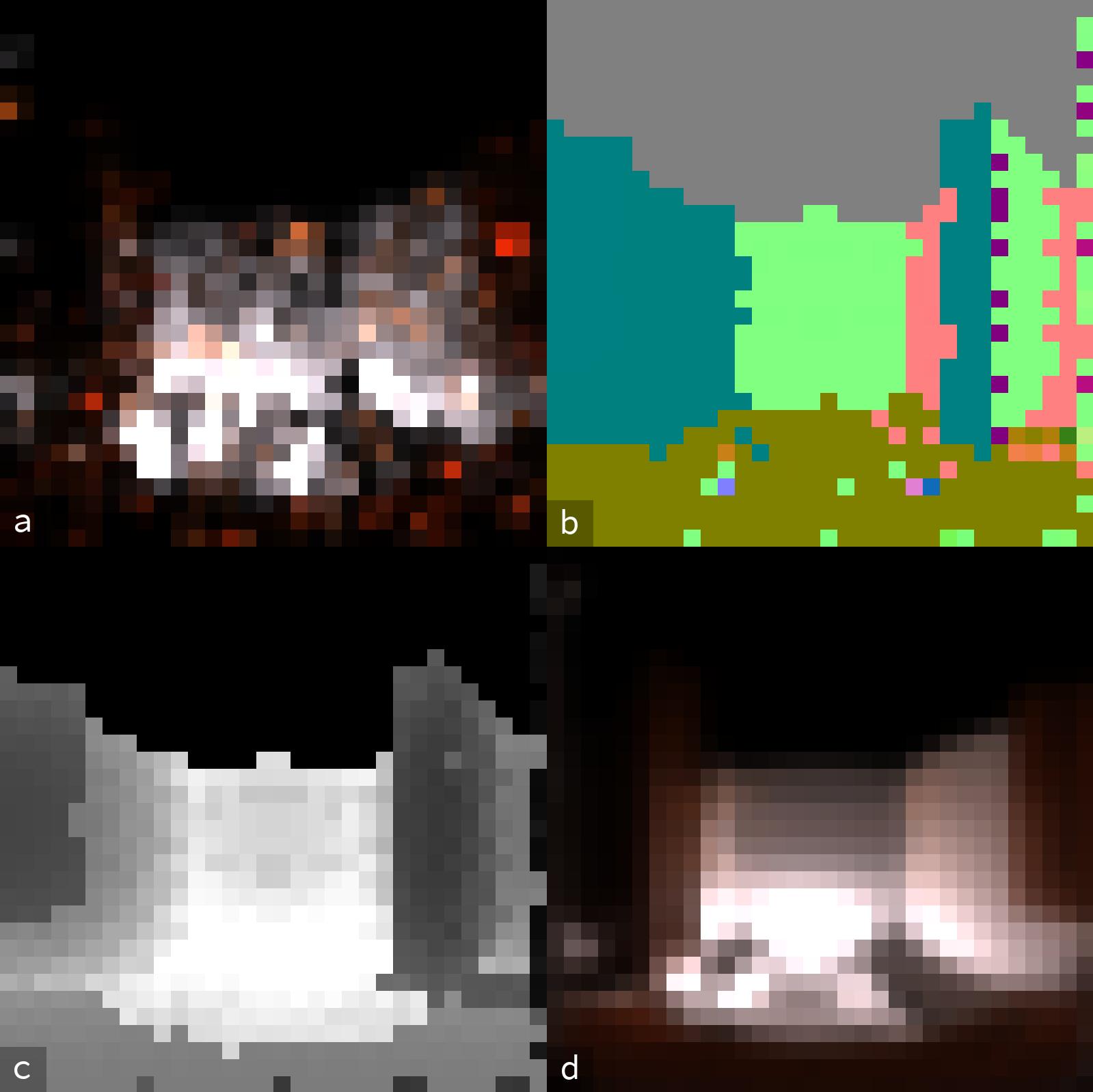}
		\caption{
			Examples of latitude-longitude hemispherical maps generated by DRC.
			This sample was generated from one intersection in the scene
			\texttt{bathroom}, part of the PBRTv3 example scenes \cite{pbrt_v3_scenes}.
			Image (a) is a path traced intensity map at 1 sample/pixel.
			Image (b) is a normals map.
			Image (c) is the distance map.
			Image (d) is the same intensity map, traced at 4096 samples/pixel, used as reference
			ground truth during training.
		}
		\label{fig:latlong_examples}
	\end{center}
\end{figure}

Each input or output layer has 32x32 floating point values. The input is composed of
7 layers: predicted radiance map (RGB), normals (XYZ), and distance (Z). The output has
3 layers: Radiance (RGB). Figure~\ref{fig:latlong_examples} shows an example for input and output maps.

The maps are encoded in hemispherical latitude-longitude equirectangular projection.
The coverage of a hemispherical map is not the full sphere, but only the
upper hemisphere visible from a scene intersection point, with the
surface normal aligned towards the Z direction of the latitude-longitude map.
The \texttt{up} vector used in the hemispherical maps is the same as the
one specified in the global scene, and it is rotated by 90 degrees towards
the global Y vector only when the surface normal also points towards Z. 
In our final model we use the L1 loss  
$L1 = \sum_{i=0}^{n} \left| y_{i} - h(x_{i}) \right|$, 
 where $y_{i}$ is a ground truth data point, and $h(x_{i})$ is a predicted value.

\begin{figure}[!htbp]
	\begin{center}
		\includegraphics[width=1.00\columnwidth]{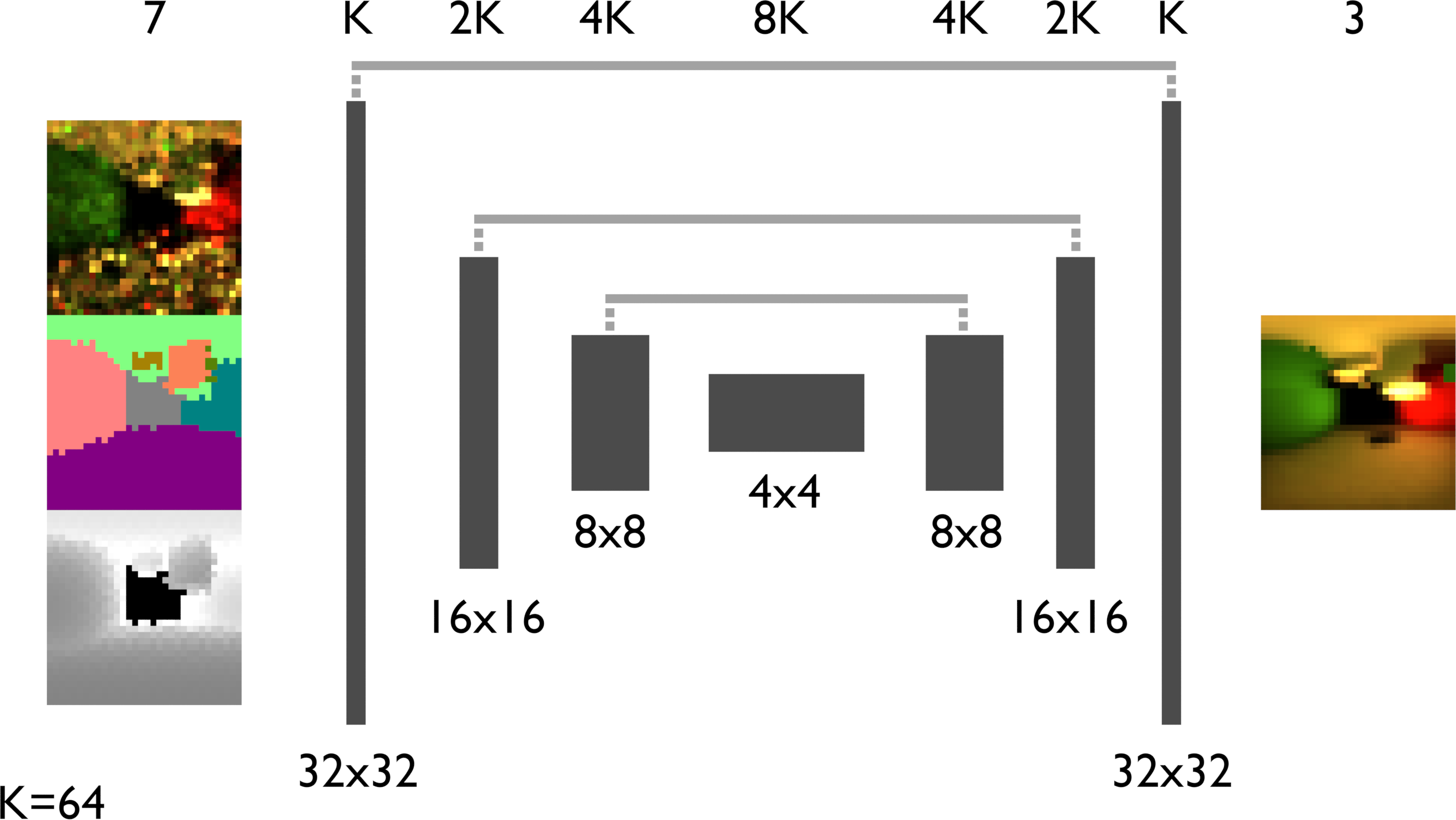}
		\caption{
			DRC Network layout
		}
		\label{fig:net_models_out_model11}
	\end{center}
\end{figure}

Our network architecture is shown in Figure~\ref{fig:net_models_out_model11} and is similar to the U-Net architecture \cite{ronneberger2015u}, a convolutional auto-encoder 
with skip connections and regular dimensionality progression. The encoder and decoder have
symmetrical structure: each encoder stage uses two 3x3 convolutions and doubles the
dimensional depth, while each decoder stage has two 3x3 deconvolutions and reduces the
depth by half.

All intermediate stages use batch normalization and LeakyReLU activation functions.
The output stage has two 3x3 deconvolutions with LeakyReLU, and a final 1x1 convolution
with ReLU activation to output the final data. Each downsampling stage
uses a 2x2 Max Pooling, and the upsampling stages use 2x2 Bilinear Upsampling.
We do not use any Dropout layer.

\section{Implementation}

\begin{figure}[!htbp]
	\begin{center}
		\includegraphics[width=1.0\columnwidth]{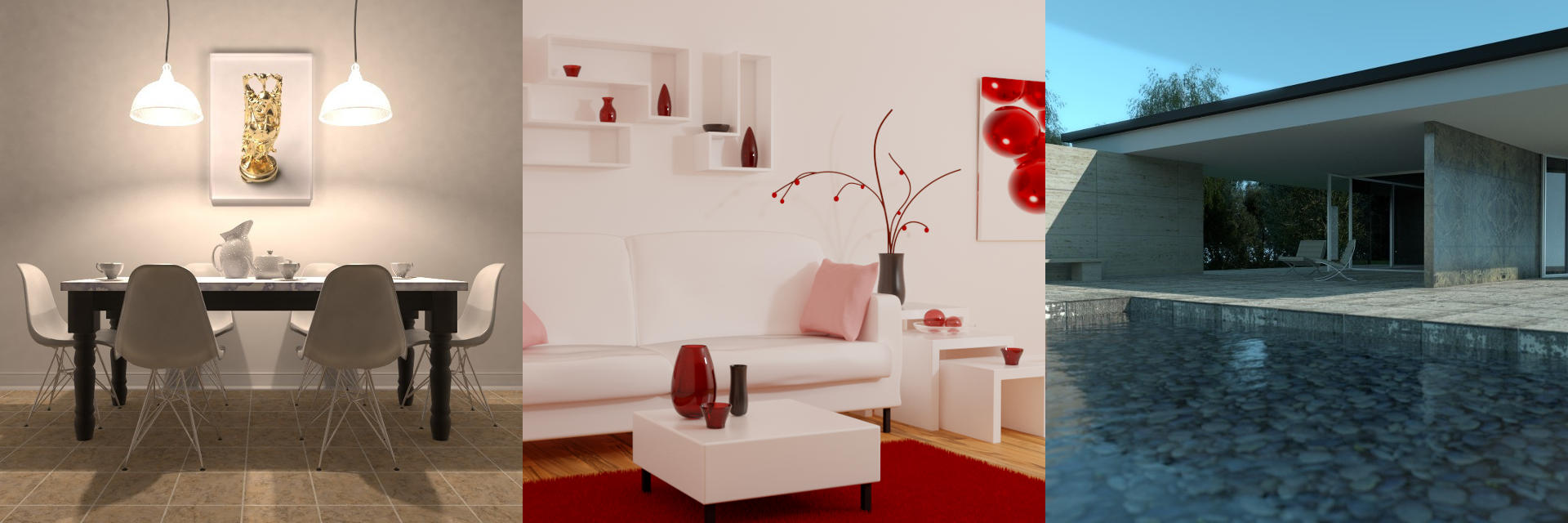}
		\caption{
			Selected scenes from the training set (cropped to fit)
		}
		\label{fig:training_example_scenes}
	\end{center}
\end{figure}

\noindent\textbf{Model training:}
A set of 43 scenes has been selected from the PBRTv3 example scenes\footnote{\url{http://pbrt.org/scenes-v3.html}},
Benedikt Bitterli's resources\cite{benedikt_scenes}
, and our own
custom created scenes. A subset of the scenes is shown in Figure \ref{fig:training_example_scenes}.
As DRC focuses on rendering indirect illumination, the selection of the training scenes has been weighted to
include mostly interior scenes with difficult illumination, and scenes with strong global illumination.

Each of the scenes has been processed to generate a set of hemispherical maps.
Similarly to the approach taken by Kalantari \cite{kalantari_machine_learning_ml_noise},
we change sampling algorithms and seeds to prevent the network from
overfitting to specific noise patterns rather than higher
level features. We use multiple samplers, including Sobol and Random,
to produce varied noise patterns.
We collected 16000 individual training examples. This dataset was sufficient in our
tests to achieve good training results.

We implemented the neural network model and training process in Python using
Pytorch\footnote{\url{http://pytorch.org/}}. Network training has been performed with GPU acceleration on an Nvidia GTX 1060 6GB card hosted by an Intel i7 4770 processor. All images and performance metrics have been acquired by using a single CPU, a Ryzen 2700. Network inference is done in DRC without any GPU acceleration.


We chose a base layer size of K=64, and a minimum hidden layer dimension of
4x4. These parameters result in 3 downsizing steps and 3 corresponding upsampling
steps.

We used the Adam optimizer included in PyTorch, with a learning rate of
$6e-5$ and implemented data augmentation in a Pytorch data loader to include combinations of rotation and flipping. We achieved good network performance after 20 minutes of training with
mini-batches containing 32 examples each.
Our training covers 3 epochs over the augmented dataset: more than half a million
individual examples.

Our full dataset is 0.5GB gzipped, and can be easily loaded into the memory of a single GPU when uncompressed during network training.
The trained model is 20MB, which can be efficiently evaluated on modern CPUs, whose caching mechanisms often exceed this size. 


\noindent\textbf{PBRT Extension:}
The DRC method and algorithm has been implemented by modifying PBRTv3 \cite{pbrt}, a C++ ray tracer
implementing many popular algorithms.

Starting from PBRTv3, we implement a new \texttt{integrator} which evaluates direct and
indirect illumination separately and communicates to the neural network to obtain 
radiance map predictions. 
PBRT-DRC can operate in two different modes: \textit{Reference} and \textit{Rendering}.

While the \textit{Rendering} mode can be used by the end-user to render a scene,
the \textit{Reference} mode computes training examples for  given  scenes. The \textit{Reference} mode automatically generates a large number of examples suitable for both
training and validation purposes.
The number of tiles can be defined, which determines the way the rendered image is subdivided into
a regular grid. Each intersection in the grid is used to shoot a single camera
ray into the scene, and the intersection found becomes the viewpoint of one
set of examples.
The high quality path tracing render's number of samples can be specified  by the user.

In the rendering process we use two independent steps to compute indirect illumination and
direct illumination.
The direct illumination pass is handled using a standard direct illumination
integrator implemented in the original PBRT software.
The indirect illumination uses the \textit{Path} integrator to collect
the neural network's input maps. The machine learning evaluation runs in a
separate Python process running Pytorch. Each rendering thread has its
own child process and communication is handled through standard Unix pipes.
Each Python evaluation process loads the saved model parameters, and reads
flattened input data from \texttt{stdin}. The predicted output is written back
to \texttt{stdout}.

\section{Evaluation and Results} \label{sec:evaluation}

DRC is evaluated in terms of its network performance, final image quality, and overall rendering performance.
We analyze the output of the Convolutional Autoencoder in Section \ref{sec:EvaluationNn},
perform an ablation study in Section \ref{sec:EvalAblationStudy} and evaluate
image quality using established metrics like SSIM and Entropy in Section \ref{sec:EvaluationFinalQuality}. SSIM is preferred in our experiments over, \emph{e.g.}, PSNR, since we are analyzing Monte Carlo rendered images~\cite{whittle2017analysis}.

\subsection{Neural network evaluation} \label{sec:EvaluationNn}

\begin{figure}[!htbp]
	\begin{center}
		\includegraphics[width=\linewidth]{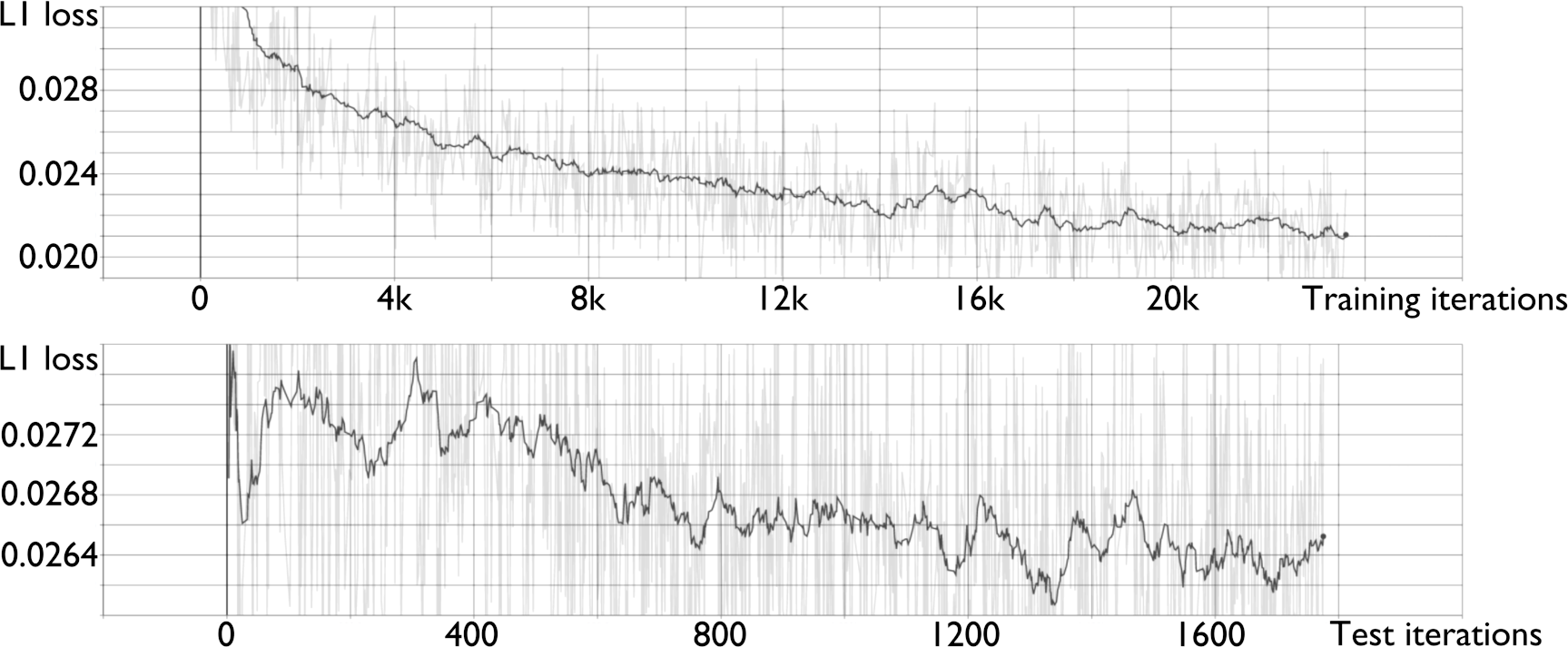}
		\caption{
			Training and Test L1 Loss in Model 11
		}
		\label{fig:prerendered_training_process_11}
	\end{center}
\end{figure}

Figure \ref{fig:prerendered_training_process_11} shows training and test
loss values during training. Our model reaches stable loss values over a 20 minutes
learning period, without overfitting to the training dataset.

The network was trained on a collection of 3D scenes, and the evaluation is conducted
on three selected scenes that were never part of the training dataset: \textit{White Room Daytime} is
part of the original PBRTv3 example scenes \cite{pbrt_v3_scenes}, \textit{Veach Ajar}
by Veach and adapted by B. Bitterli~\cite{benedikt_scenes}, and \textit{Mbed1},
custom created for the purpose of this evaluation.

The difference in amount and depth of indirect illumination paths across the test
scenes allows us to verify that our network is able to deal with different input 
data quality seamlessly, and that it is able to extrapolate and adapt to geometry not
encountered during training.

As with the training set generation, we obtain input maps and ground truth path traced
radiance maps for several viewpoints located at primary surface intersection points.
While the PBRT-DRC implementation re-normalizes the output images to match the expected
intensity level of the entire radiance map, in our tests we compare directly using
the final intensities obtained after the upstream processing to provide an unbiased
view of the network performance.


\begin{figure*}[!htpb]
	\vspace{-1.0cm}
	\begin{center}
		\includegraphics[width=\textwidth]{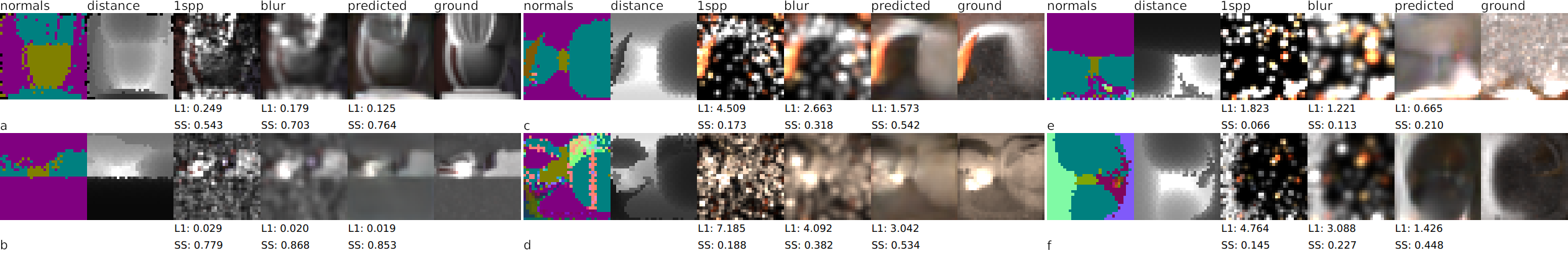}
		\caption{
			A random selection from the test scenes \textit{Veach Ajar},
			\textit{Custom Mbed}, \textit{White Room Daytime}. The normals are mapped via $n_i = ((n_i * 0.5) + 0.5)$ and then scaled to RGB space. 
			All examples show excellent noise removal when comparing \textit{1spp}
			to \textit{predicted}, with the predictions occasionally presenting
			less noise than the reference ground truth.
			In e) there is some color inaccuracy, with the predicted image showing some
			slight tint that is not present in the ground truth.
			e) and f), part of \textit{Veach Ajar}, are very difficult due to the very sparse
			sampling of the input intensity map. They show excellent noise removal,
			producing less noise than the ground truth.
			c) shows a slight shape distortion on the orange part, and a), b) and d) are less sharp
			than the ground truth, but still considerably more usable than the Gaussian blur versions.
		}
		\label{fig:prerendered_nnSelection2}
		\vspace{-0.8cm}
	\end{center}
\end{figure*}

Figure~\ref{fig:prerendered_nnSelection2} shows a random selection of radiance maps
predicted using our neural network model. The first three columns show the
inputs of the network: normals are tonemapped from XYZ to RGB via $n_i = ((n_i * 0.5) + 0.5)*255$, distances from the camera and 1 sample/pixel path tracing radiance maps.
The result is compared to a simple Gaussian blur and
a reference path traced ground truth, rendered at 1024 samples per pixel.
As many of the initial path tracer outputs comprise mostly of black pixels and a few
very bright ones, standard image denoising filters such as Bilateral and Median Filter do not perform well. The Gaussian blur, although yielding softer outputs,
is more capable at filling the black gaps.

The Gaussian blur uses a one standard deviation filter width, chosen for its
good compromise between smoothing power and blurriness. A smaller radius would
not be able to remove much noise, while a larger radius would yield considerably
blurrier results compared to either ground truth or network predicted output.
Due to the very sparse samples generated by the path tracer, a simple Gaussian
blur is a significant improvement over the 1spp (1 sample per pixel) intensity map.

The examples show that the network predicted output can consistently outperform
the Gaussian blur by producing sharper images, preserving more detail, and
removing more noise. We can observe that:

\begin{itemize}
	
	\item The predicted result is generally sharper than the Gaussian blur.
	This confirms that we would not be able to use a larger radius
	for the Gaussian blur for difficult cases without losing more details.
	
	\item The network is able to reconstruct some details that are missing
	from the 1spp intensity map, although it can cause artifacts.
	
	\item The network has effectively learned to use adaptive blurring filters.
	In simple cases it behaves very similarly to a planar Gaussian blur, but in
	the general case it is able to adapt to different sampling densities more
	effectively.
	
	\item The network eliminates all visible noise, and can produce images
	that look smoother with less noise when compared to the high quality
	path traced reference images. 
	This shows the ability of the network to extrapolate
	from the training dataset, which still contains a small amount of noise in
	its ground truth examples. This is in line with findings from Noise2Noise \cite{lehtinen2018noise2noise}.
	
\end{itemize}

To verify our claims, we evaluate similarity metrics on a large number of radiance evaluation points
collected from the testing scenes, totaling over 1000 distinct testing examples
that were not seen during training. For each example set of images, we compute the L1 absolute difference
between predicted and reference images 
\begin{equation}
L1 = \frac{1}{n} \sum_{i=0}^{n} \left| p_{i} - g_{i} \right|, 
\end{equation}

\begin{figure*}[!htbp]
	\begin{center}
		\subfloat[][]{\includegraphics[width=0.47\textwidth]{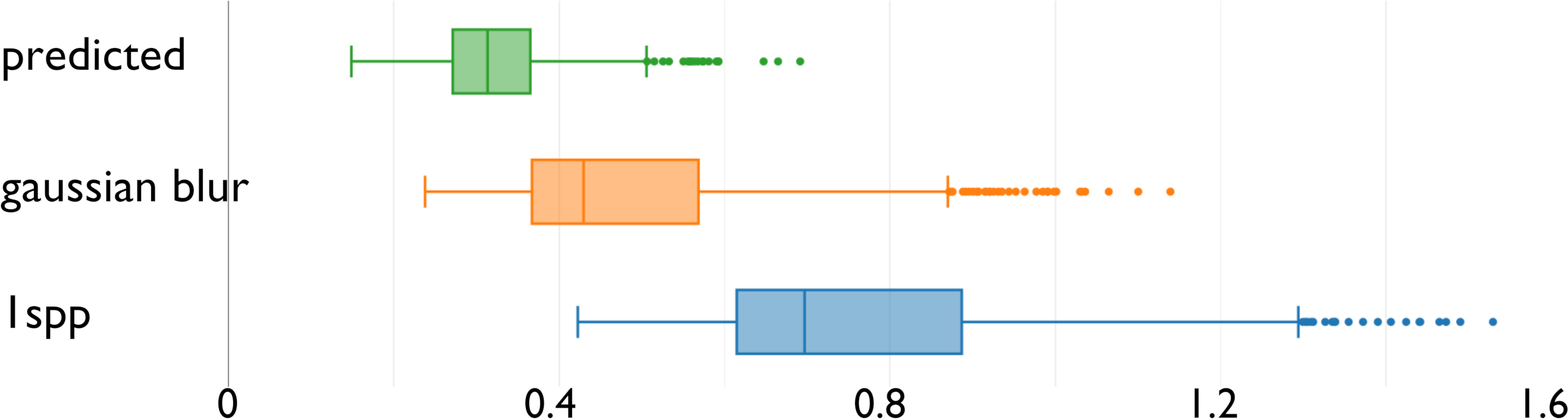}}
		\subfloat[][]{\includegraphics[width=0.49\textwidth]{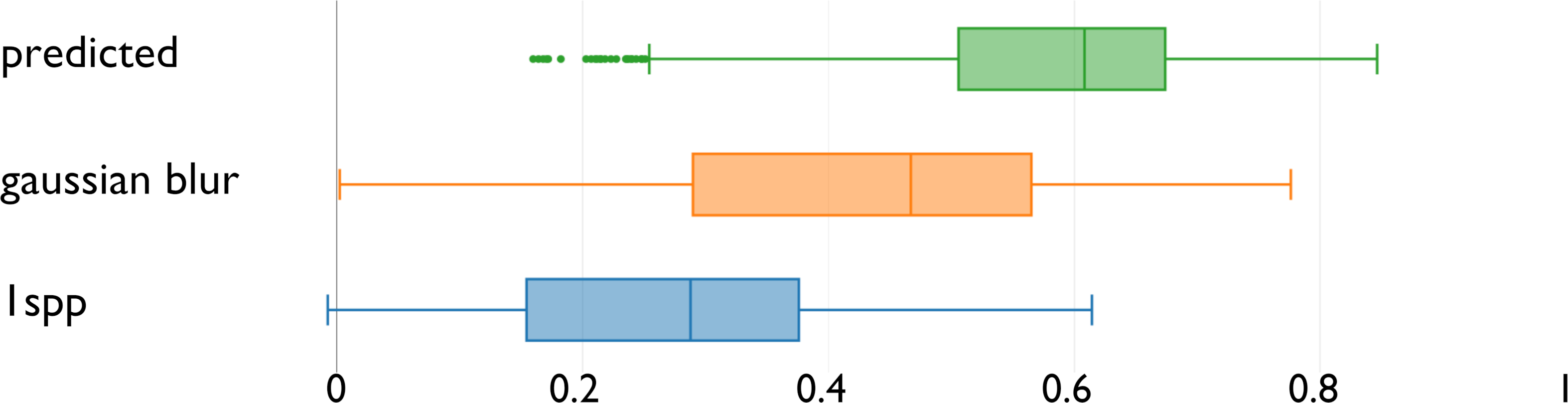}}
		\caption{
			(a) L1 absolute difference, (b) Structural Similarity
		}
		\label{fig:prerendered_boxl1}
	\end{center}
\end{figure*}
where $p_{i}$ is a predicted pixel value, $g_{i}$ is a reference pixel value.
Figure \ref{fig:prerendered_boxl1} shows L1 and Structural Similarity Index \cite{wangSsim2004}. The Gaussian blur does considerably
increase the quality of the radiance maps compared
to using raw 1spp images. The structural similarity increases, as the blur
fills many of the black gaps present in the path traced map.

The network output considerably outperforms the Gaussian blur in both L1 and
structural similarity metrics. We confirm that for both metrics the samples 
belong to different probability distributions by computing the Kruskal-Wallis 
null hypothesis testing \cite{kruskal1952use} p values for L1 and
Structural Similarity. We verified the p values to be less than 0.00001.

\subsection{Neural network ablation study} \label{sec:EvalAblationStudy}

The ablation study aims to verify  hypotheses about the data that the neural network uses to
produce accurate predictions. Training is conducted using the entire dataset of intensity, normals, and distance maps, and repeated with the normals, distance, or both components
removed from the network's inputs. The structure of the network is left unchanged,
and the ablation is implemented by setting the target components to be zero arrays.
The evaluation  is conducted in a similar way, by setting the ablated channels
to zero arrays.

\begin{figure}[!htbp]
	\begin{center}
		\includegraphics[width=0.5\textwidth]{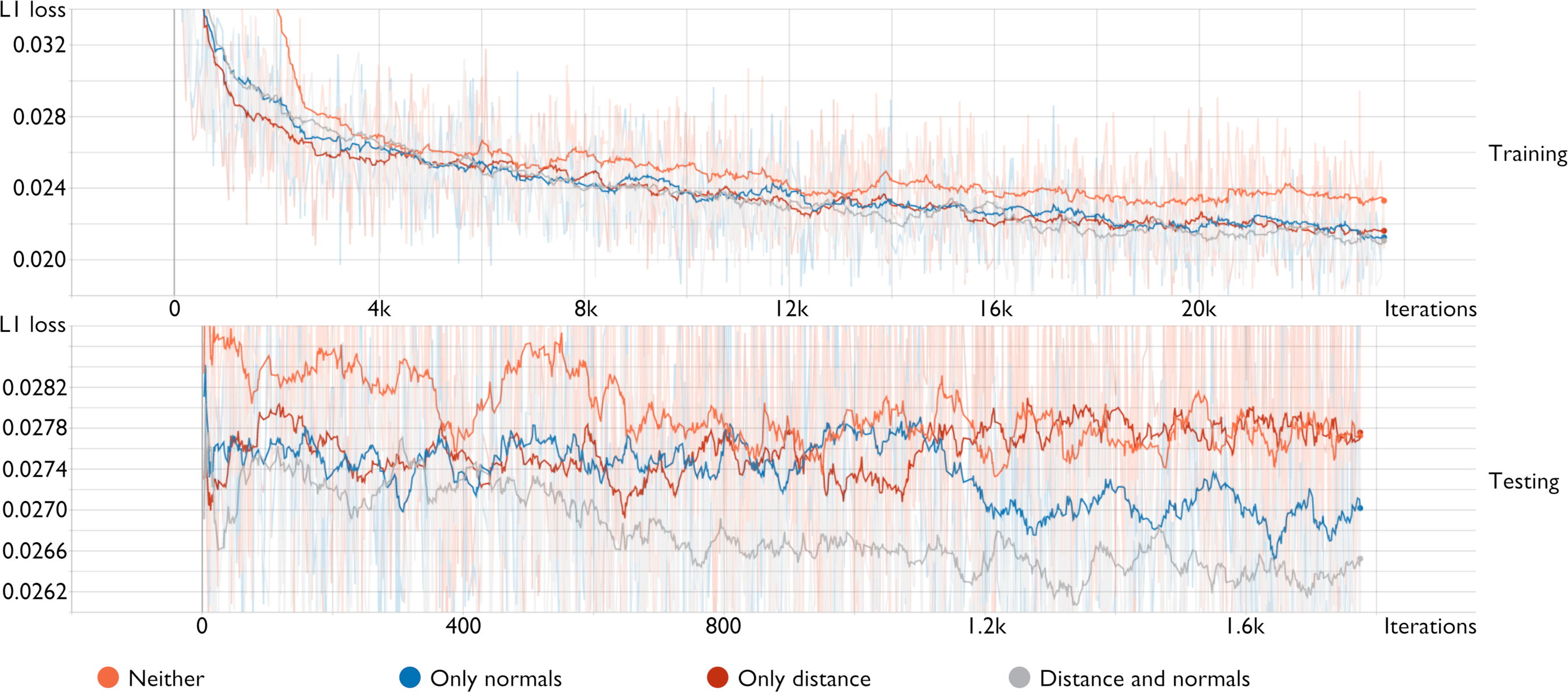}
		\caption{
			Training loss ablation testing. Removing information from the input data
			slows down convergence of the loss values. The effect is visible
			on the training set, but becomes more evident on the test set.
		}
		\label{fig:ablationTestOutTensorboardJpg}
	\end{center}
	\vspace{-0.8cm}
\end{figure}
Figure \ref{fig:ablationTestOutTensorboardJpg} shows loss values during training and
testing using the four different configurations. The use of only first hit path tracing maps
performs better than a simple blur, but is improved by the use of
normal maps. Using distances without normals does not yield better performance,
but the combination of all three inputs outperforms the other
configurations.

\begin{figure}[!b]
	\begin{center}
		\includegraphics[width=0.45\textwidth]{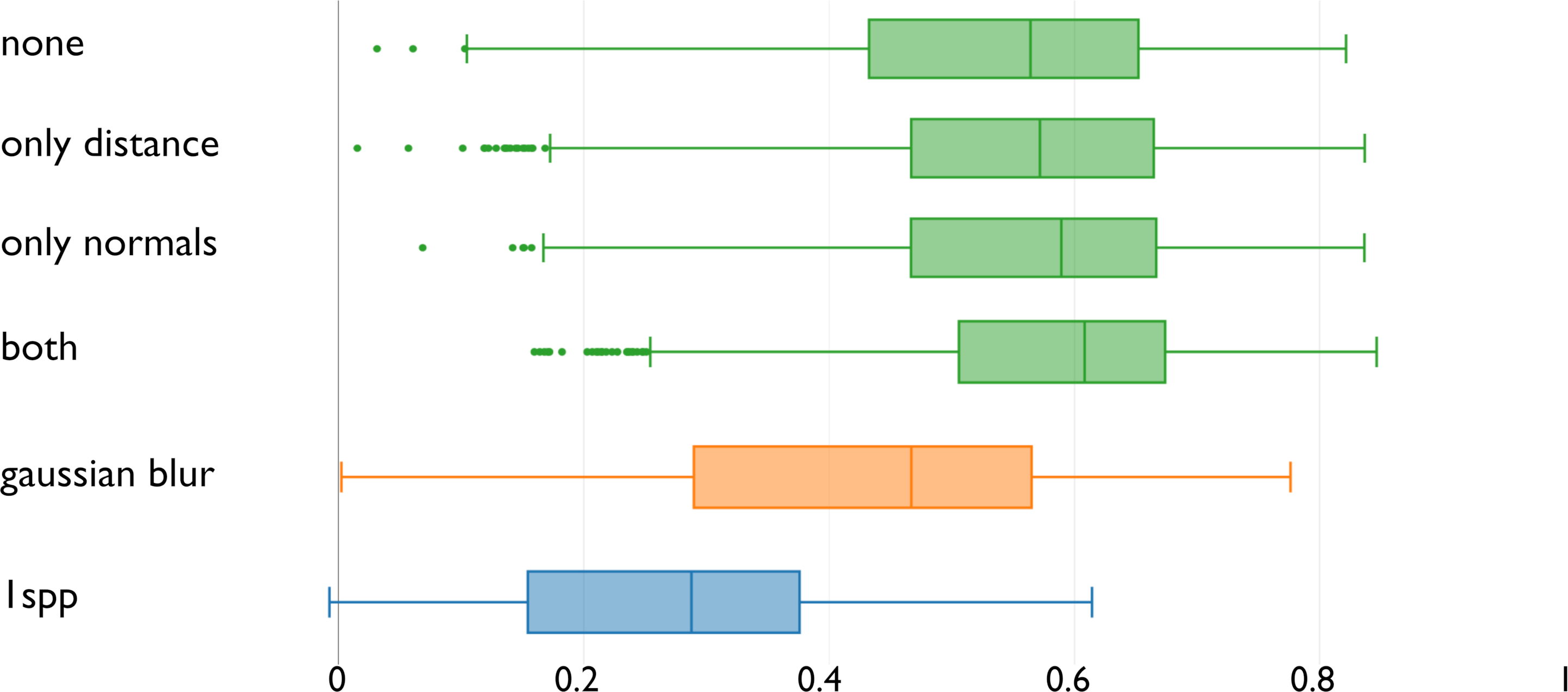}
		\caption{
			The box plots show structural similarity values in our ablation testing.
			None: predicted results with the network trained on intensity maps only.
			Only distance: network trained with intensity and distance maps.
			Only normals: network trained with intensity and normal maps.
			Both: network trained with intensity, distance and normal maps.
			Gaussian blur: Gaussian blur applied to the input intensity map.
			1spp: input intensity map path traced at 1 sample per pixel.
		}
		\label{fig:ablationTestBoxOutSsimAblationJpg}
	\end{center}
\end{figure}
The Structural Similarity Index test on the entire test set in Figure
\ref{fig:ablationTestBoxOutSsimAblationJpg} confirms that the addition of normal
or distance maps helps the network to predict more accurate results, with the
normals being slightly more effective than distances alone. The combination of both
additional layers further increases accuracy, and in particular improves the
lower quartile notably.

\subsection{Final rendering evaluation} \label{sec:EvaluationFinalQuality}

The evaluation of the final images is based on both qualitative and quantitative metrics.
We render each of our test scenes at different levels of quality exclusively on a CPU (Ryzen 2700 without GPU acceleration), and present observations,
statistics and comparisons.


DRC is not based on a per-pixel quality
setting, but on two distinct settings: the number of samples per pixels for the
direct illumination integrator, and the number of tasks to be processed
in the indirect illumination pass. 
A \textit{task} is a single computational job in the indirect illumination pass 
that corresponds to a rectangular region of the rendered image. A task is responsible
for collecting the input maps in the region, evaluating them through the neural
network, and adding the final light contributions to the image. DRC heuristically
splits indirect illumination rendering into tasks to distribute the computational
load onto multiple CPU cores.

\begin{table}[ht] 
	\caption{
		Quality and noise metrics on the Bathroom Green scene~\cite{cenobi_bathroom_green} with an increasing number of indirect illumination tasks. We compare DRC to BDPT~\cite{lafortune1993bi} integration numerically.  We evaluate \textit{Render time} as total rendering time in seconds; 
		\textit{SSIM}: the Structural Similarity Index; and the \textit{PNG File Size} as a measure of the amount of noise in the final image. Better or equal values are highlighted in bold.
	}
	\begin{tabular}{rrrrrrrr}
		\toprule
		& \multicolumn{3}{c}{\textbf{BDPT}}    & \multicolumn{4}{c}{\textbf{DRC}}\\ \cmidrule(r){2-4} \cmidrule(r){6-8}
		\multicolumn{1}{c}{Tasks} & 
		\multicolumn{1}{c}{\begin{tabular}[c]{@{}c@{}}Time \\ (s)\end{tabular}} & \multicolumn{1}{c}{SSIM} & 
		\multicolumn{1}{c}{\begin{tabular}[c]{@{}c@{}}PNG \\ (KB)\end{tabular}} & 
		\multicolumn{1}{c}{} &
		\multicolumn{1}{c}{\begin{tabular}[c]{@{}c@{}}Time \\ (s)\end{tabular}} & \multicolumn{1}{c}{SSIM} & 
		\multicolumn{1}{c}{\begin{tabular}[c]{@{}c@{}}PNG \\ (KB)\end{tabular}} \\ \midrule
		1& \textbf{3} & 0.48 & 2497 && 36& \textbf{0.77} & \textbf{2482} \\ 
		2& \textbf{7} & 0.57 & 3094 && 36& \textbf{0.84} & \textbf{2691} \\
		4& \textbf{12} & 0.66 & 3176 && 38& \textbf{0.87} & \textbf{2549} \\ 
		8& \textbf{23}& 0.75 & 3050 && 39& \textbf{0.88} & \textbf{2486} \\ 
		16    & 44& 0.82 & 2885 && \textbf{43}& \textbf{0.89} & \textbf{2412} \\ 
		32    & 95& 0.87 & 2710 && \textbf{59}& \textbf{0.89} & \textbf{2354} \\ 
		64    & 181  & \textbf{0.90} &2549 && \textbf{88}& \textbf{0.90} & \textbf{2313} \\ 
		128   & 351  & \textbf{0.92} & 2392 && \textbf{160}  & 0.91 & \textbf{2278} \\ 
		256   & 693  & \textbf{0.93} & 2258 && \textbf{322}  & 0.91 & \textbf{2250} \\ \bottomrule
	\end{tabular}
	\label{Tab:bathroom_green}
\end{table}

We chose a target quality setting by rendering the scene at a fixed number off direct lighting passes,
while changing the number of indirect lighting tasks.
For this comparison, we use both DRC and the bidirectional path tracer integrator from PBRTv3 in CPU-only mode, without using any dedicated accelerator
to evaluate the neural network. We used a PC with a Ryzen 2700 and 16GB of RAM.
We measure the output's Structural Similarity compared to a high quality path traced image rendered at 2048 samples/pixel,
and the PNG file size as an indication of the amount of noise.
Table~\ref{Tab:bathroom_green} shows that DRC's overhead on render time is very small until 16 indirect tasks,
with Structural Similarity values increasing. By using more than 16 indirect tasks, quality improves only
marginally, although rendering takes a much longer time. As DRC is a biased method, Structural Similarity does
not converge towards 1, and any artifacts present in the output need to be judged subjectively.
The amount of noise continues decreasing as the indirect tasks increase, although the noise in the output
image is mostly produced by the direct illumination passes.
We therefore use 16 indirect samples for our comparisons, and 8 direct lighting samples as it is a good compromise
between noise and speed.

We compare the DRC result at 8 direct and 16 indirect passes, which took 43 seconds, with path tracing at 16 samples/pixel completed in a similar amount of time, 46 seconds, and
bidirectional path tracing completed in 44 seconds.
A reference path traced image at 2048 samples/pixel is also included (99m 30s).

In the comparison, we also include results post processed by the Intel\textsuperscript{\textregistered}  Open Image Denoiser~\cite{Intel2020}, a production-quality image denoiser built to be used with path traced images. We used the Open Image Denoiser with all settings at default, in
LDR mode.

\begin{figure*}[t] 
	\begin{center}
		\includegraphics[width=1.0\textwidth]{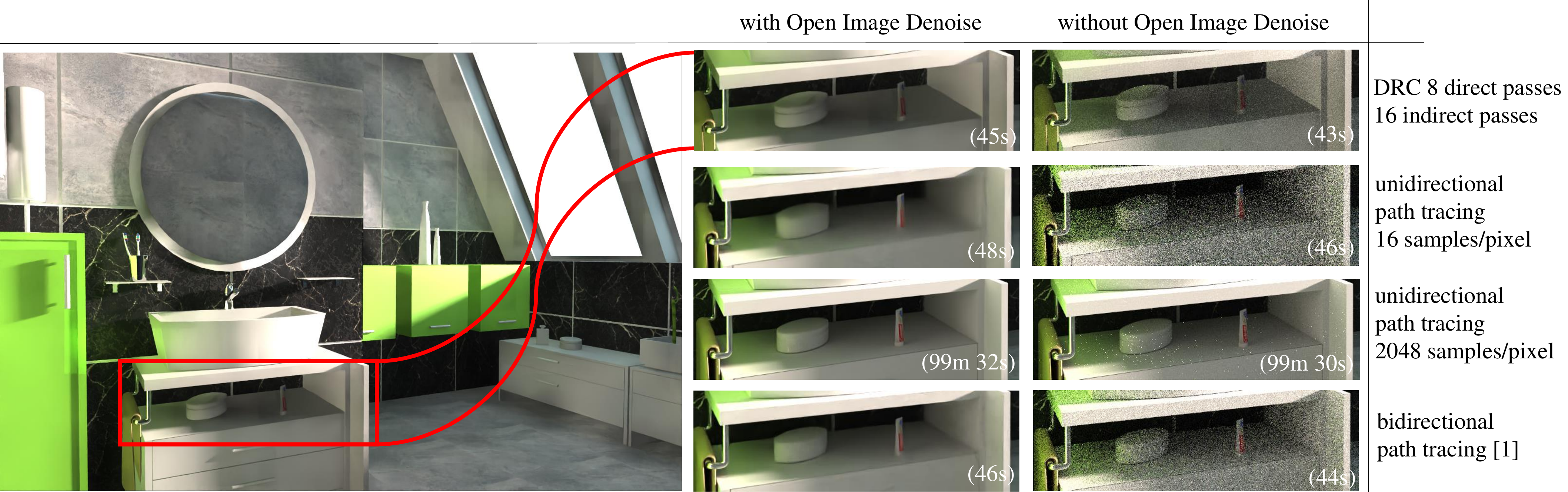}
		\caption{
			Comparison between DRC, path tracing, and Bi-Directional path tracing on Bathroom Green~\cite{cenobi_bathroom_green}.
			path tracing at 2048 samples/pixel should be considered the reference image. We have adjusted render
			quality settings for other methods to yield similar computation times. Note that the camera angle and some materials and textures are not identical to~\cite{cenobi_bathroom_green} since we had to manually convert the scene to PBRT. 
		}
		\label{fig:renderComparisonImagesBathroomGreen}
	\end{center}
\end{figure*}

Figure~\ref{fig:renderComparisonImagesBathroomGreen} shows a comparison between
path tracing, DRC, bidirectional path tracing and each of them after applying the Open Image Denoiser. 
The DRC output appears  less noisy
than an image from path tracing after the same amount of rendering time. 
The generated global illumination effects  are visually improved in the DRC output, including the 
green bleeding onto the slightly-glossy sink, and the ambient occlusion in the darker areas.
The crops show that DRC is capable of a higher level of detail, especially in poorly illuminated areas.
The Open Image Denoiser framework works well for removing all noise, although the filtered path tracing output at 16 samples/pixel
has a notable loss of detail in the area below the sink, while the filtered DRC output retains more details.

It is important to point out that the noise that is still visible in the DRC output
is not caused by our method, but by the re-integration of the direct illumination
layer.

\begin{figure*}[!htbp]
	\centering
	\subfloat[][]{\includegraphics[width=0.19\textwidth]{prerendered/glossybox/five-direct.jpg}} \hfill
	\subfloat[][]{\includegraphics[width=0.19\textwidth]{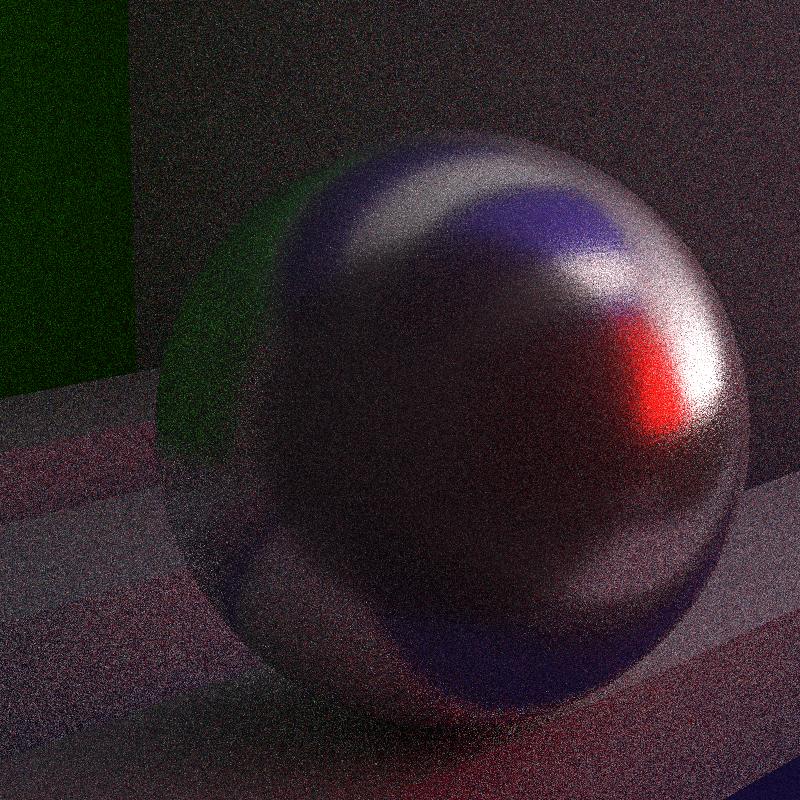}} \hfill
	\subfloat[][]{\includegraphics[width=0.19\textwidth]{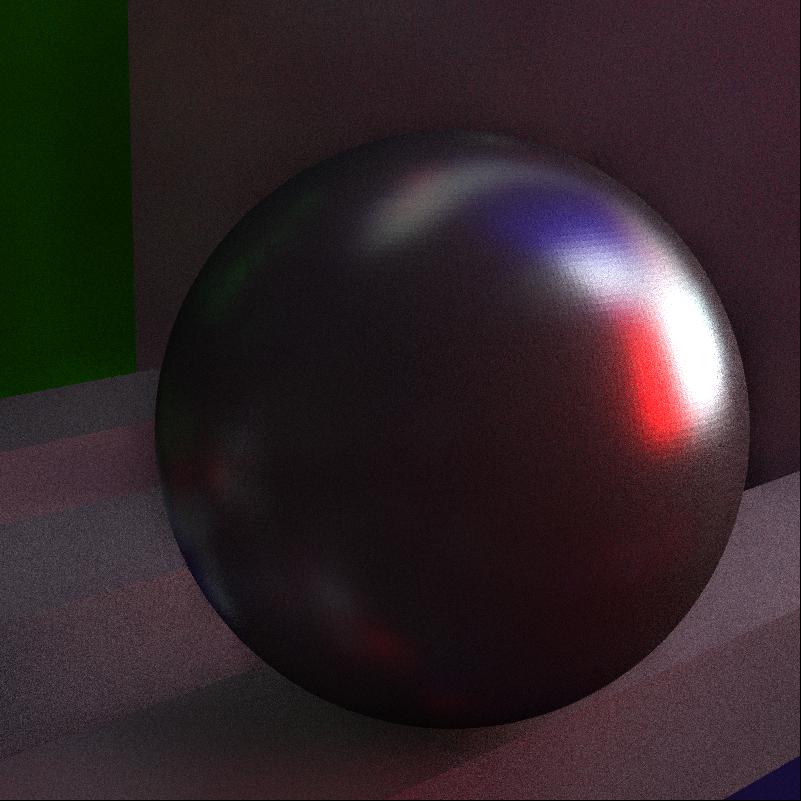}} \hfill
	\subfloat[][]{\includegraphics[width=0.19\textwidth]{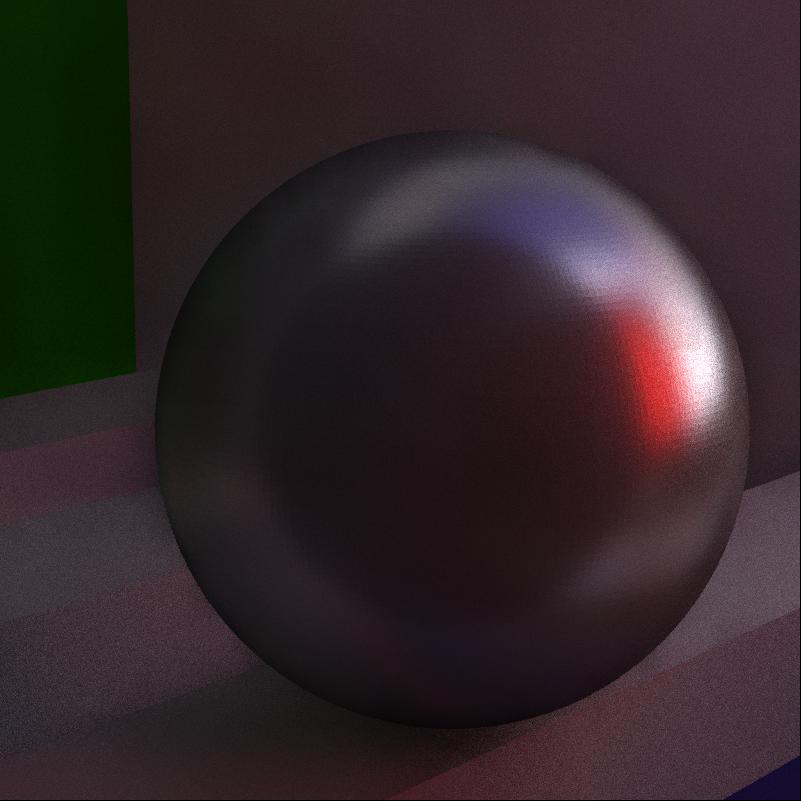}} \hfill
	\subfloat[][]{\includegraphics[width=0.19\textwidth]{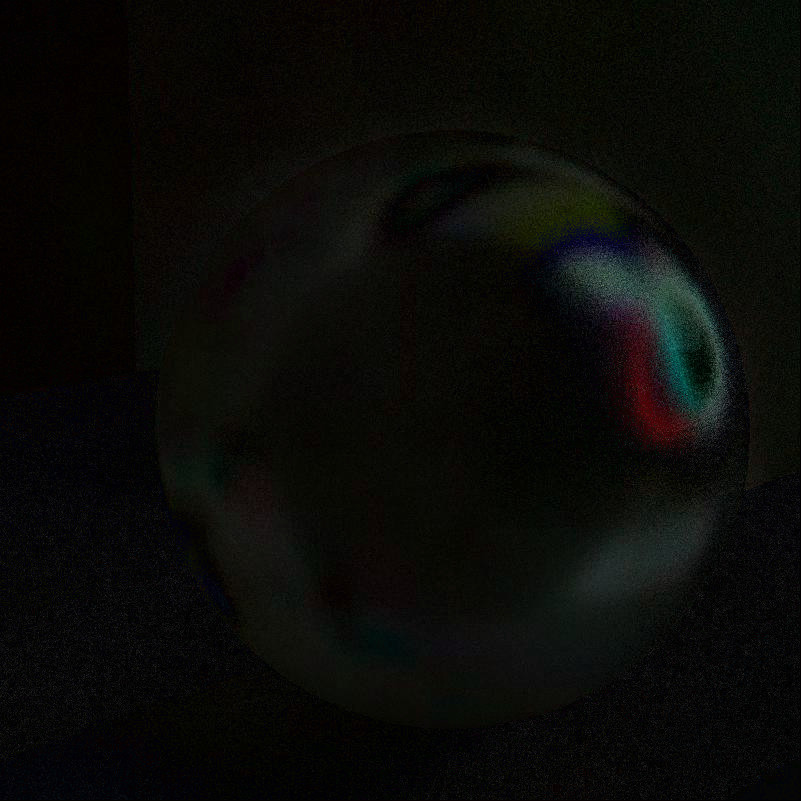}}
	\caption{ A direct comparison between 
		(a) Direct illumination, 
		(b) Path tracing, 
		(c) DRC Gaussian blur, 
		(d) DRC Neural Network, and (e) the difference between between (c) and (d). The baseline method (c) misses several details on the lower left part of the sphere that have been correctly predicted in (d) when compared to (b). MSE: (c), (b) = 500.31; (d), (b) = 384.81. SSIM: (c), (b) = 0.21; (d), (b) = 0.23.
	}
	\label{fig:prerenderedGlossybox}
\end{figure*}

\noindent\textbf{Glossy and specular materials: }
One of the key properties of DRC is its ability to work with a wide range of materials,
including glossy surfaces.  Global illumination reflections heavily depend on the
quality of the radiance maps.
The neural network evaluation (Section~\ref{sec:EvaluationNn})
shows that a Gaussian blur applied to the intensity maps offers a good level of similarity
compared to the ground truth. Figure~\ref{fig:prerenderedGlossybox} shows
a sphere with a highly glossy surface under difficult lighting conditions. The direct illumination
integrator in  Figure~\ref{fig:prerenderedGlossybox}(a) is almost completely black, confirming that all reflections are indirect.
Blurred radiance maps may be sufficient to handle materials with a low level of glossiness or on low importance scene objects. The neural network in DRC brings improvements in the rendering quality
of shinier surfaces.
For example, blurred radiance maps do not have sufficient resolution
in the dark areas (compare Figure~\ref{fig:prerenderedGlossybox}(c) and (d)) and the neural network produces more stable and detailed output so that reflected shapes are better preserved. 
In contrast to denoising and blur filters, DRC can further be tuned by increasing the size of the predicted radiance maps, which directly reduces resolution-induced blur in the final result. 
The computational overhead of (d) compared to (c) is $~40\%$, which can be improved in future extensions by adaptively choosing a radiance map enhancement strategy depending on the material properties.

Specular highlights, reflections and transmission materials are handled by
following the original ray, and evaluating the radiance maps after the bounce.
The threshold between the first and secondary ray bounce in DRC has therefore slightly
different semantics from most ray tracers
as perfect mirrors and transmission rays do not increase the depth counter.
These specular rays are also ignored in the depth count of the direct illumination
pass of DRC to provide a correct complementary integrator.
In the details of the Bathroom Green scene
in Figure \ref{fig:renderComparisonImagesBathroomGreen},
the metallic objects preserve detailed reflections and realistic structural details are predicted. 


\begin{figure}[!htbp]
	\captionsetup[subfigure]{justification=centering}
	\centering
	\subfloat[][]{\includegraphics[width=0.49\columnwidth]{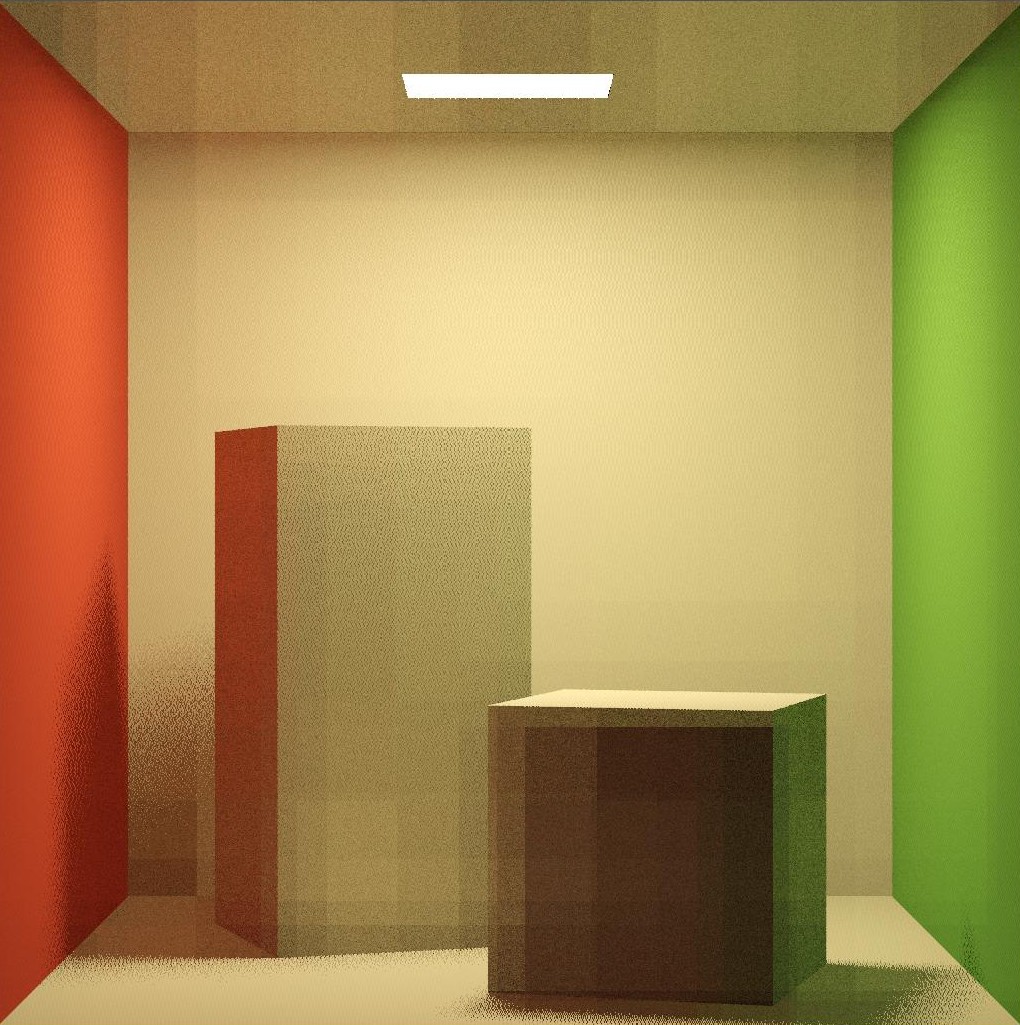}} \hfill
	\subfloat[][]{\includegraphics[width=0.49\columnwidth]{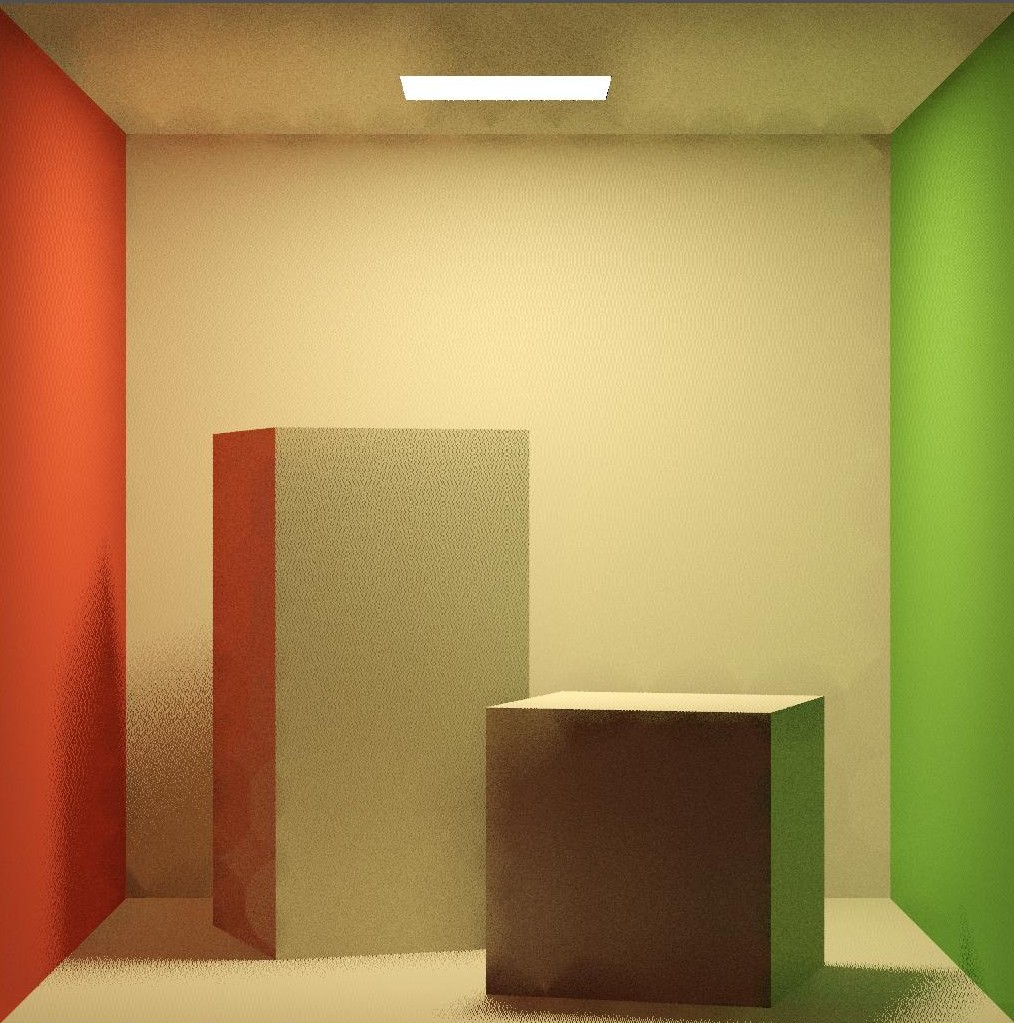}}
	\caption{
		Comparison between simple distance-based interpolation (a)
		and interpolation
		based on both distance and surface normals (b)
		Using surface normals improves smoothness across surfaces, and
		removes large artifacts and light bleeding on the edges of the
		cubes.
	}
	\label{fig:interpolationArtifacts}
\end{figure}

\noindent\textbf{Interpolation:}
Figure \ref{fig:interpolationArtifacts} shows two images rendered at a limited
number of tasks. The simple interpolation strategy linearly interpolates using
pixel distances on the image plane. Light bleeding is clearly visible on the objects. 
This image also shows hard edges, artifacts resulting from individual threads rendering
different tiles of the image, and the lack of interpolation among adjacent tiles.
We resolve some major artifacts by including
surface normals in the weighting system (Section~\ref{sec:progressRefinInter}),
and by improving the way tiles are overlapped in the image.
Some artifacts remain near the edges
due to the very low number of samples, while the overall image looks smooth.

Increasing the number of indirect illumination passes in DRC only affects computational time. Memory usage remains constant, as each interpolation pass is independent and does not require any data from previous passes.

\section{Discussion}

The DRC method offers a new biased rendering method for global illumination capable of producing
high quality results efficiently, while supporting a wide range of
material types and scene compositions.

Compared to other methods, DRC does not require any form of pre-computation on the target
scene, and accelerates the computation of global illumination at a virtually unbounded number
of indirect bounces at the cost of a small amount of bias introduced.
The typical use case scenario is rendering an interior 3D scene where indirect light is significant. Most interior scenes have strong indirect lighting, making light paths computationally intensive.

Faster estimation of indirect light through a neural network can drastically reduce the amount of noise present in the scene,
and the typical low-frequency variation of indirect illumination easily masks away minor artifacts and interpolation imprecisions
that originate from the machine learning subsystem. Thanks to the preservation of a standard direct illumination pass in the final image,
most of the details and textures present in the scene are preserved in the final output.

DRC splits the processing of direct and indirect illumination, allowing the user to better tune the amount of computation to be used for each component.

\subsection{Limitations}
Even though the progressive refinement algorithm permits the user to set an arbitrary number
of passes, DRC remains a biased approach to global illumination. The neural network,
although very accurate in typical scenarios, can produce artifacts, inaccurate results,
and loss of detail, causing hard to correct errors.

A second important limiting factor is the fixed resolution of the radiance maps, set for DRC at
a default of 32x32 pixels. Small details are inevitably lost.

These two limitations become particularly visible when computing globally illuminated
glossy reflections, which appear less defined, although less noisy, than the path traced
reference.

Temporal stability is another limitation of DRC. The rendering tiles
can have large variations from one sample to another, and while being perceived as smooth
transitions in a single frame, animations would display them as low frequency flickering
of large areas in the scene. The lack of temporal stability indicates that DRC would
not be adequate for animation rendering. To solve the temporal stability issue, 
a persistent caching system for radiance maps
can be implemented, allowing animations to be rendered at faster rates and with less
temporal noise caused by flickering of radiance maps.
Precomputing and caching radiance maps would however not be sufficient to achieve temporal stability in
scenes with dynamic objects.

Due to the low resolution of the indirect radiance maps, DRC is not capable of
producing accurate caustics, which require a much denser sampling of indirect
light.




The progressive refinement algorithm used in DRC requires no preprocessing and
very little additional memory to run. However in some scenarios predetermined
and optimized radiance interpolation points can be a more effective solution,
and implementing the possibility to choose between the two strategies would
be a useful future extension.

\section{Conclusion}

We have presented \textit{Deep Radiance Caching} -- DRC, a novel ray tracing method
that uses Convolutional Autoencoders to obtain high performance and accurate global illumination
effects. We show that our method achieves competitive performance on a CPU, while being
able to produce higher quality images than same-time path tracing, with a significantly
smaller amount of noise. Our method supports a wide range of material types
and does not need offline pre-computation or per-scene training. If required, both, the used deep learning network and the path tracing core map naturally to parallel hardware like GPUs, which has been shown to achieve real-time performance for ray-tracing~\cite{parker2010optix,goodfellow2016deep} and denoising ~\cite{mara2017efficient}. Recent developments like Nvidia's RTX technology~\cite{burgess2020rtx} opens up further avenues to exploit deep neural networks deeper in image generation algorithms. 

\noindent\textbf{Resources:}
A website including plugin for Blender and a source code repository on github are available and will be added here after the anonymous reviewing phase.

\noindent\textbf{Acknowledgements:} 
This work has been kindly supported by Intel\textsuperscript{\textregistered} and Nvidia.

\bibliographystyle{cag-num-names}
\bibliography{egbibsample}

\begin{thebibliography}{47}
\providecommand{\natexlab}[1]{#1}
\providecommand{\url}[1]{\texttt{#1}}
\providecommand{\href}[2]{#2}
\providecommand{\path}[1]{#1}
\providecommand{\eprint}[1]{\href{http://arxiv.org/abs/#1}{\path{#1}}}
\providecommand{\DOIprefix}{doi:}
\providecommand{\ArXivprefix}{arXiv:}
\providecommand{\URLprefix}{URL: }
\providecommand{\Pubmedprefix}{pmid:}
\providecommand{\doi}[1]{\href{http://dx.doi.org/#1}{\path{#1}}}
\providecommand{\Pubmed}[1]{\href{pmid:#1}{\path{#1}}}
\providecommand{\BIBand}{and}
\providecommand{\bibinfo}[2]{#2}
\ifx\xfnm\undefined \def\xfnm[#1]{\unskip,\space#1}\fi
\bibitem[{Lafortune and Willems(1993)}]{lafortune1993bi}
\bibinfo{author}{Lafortune\xfnm[ EP]}, \bibinfo{author}{Willems\xfnm[ Y]}.
\newblock \bibinfo{title}{Bi-directional path tracing}.
\newblock In: \bibinfo{booktitle}{Compugraphics' 93}. \bibinfo{year}{1993}, p.
  \bibinfo{pages}{145--153}.
\bibitem[{Veach and Guibas(1997)}]{veach1997metropolis}
\bibinfo{author}{Veach\xfnm[ E]}, \bibinfo{author}{Guibas\xfnm[ LJ]}.
\newblock \bibinfo{title}{Metropolis light transport}.
\newblock In: \bibinfo{booktitle}{Proceedings of the 24th annual conference on
  Computer graphics and interactive techniques}. \bibinfo{organization}{ACM
  Press/Addison-Wesley Publishing Co.}; \bibinfo{year}{1997}, p.
  \bibinfo{pages}{65--76}.
\bibitem[{Pharr et~al.(2018)Pharr, Jakob and Humphreys}]{pbrt_v3_scenes}
\bibinfo{author}{Pharr\xfnm[ M]}, \bibinfo{author}{Jakob\xfnm[ W]},
  \bibinfo{author}{Humphreys\xfnm[ G]}.
\newblock \bibinfo{title}{Scenes for pbrt-v3}.
\newblock \bibinfo{howpublished}{\url{http://pbrt.org/scenes-v3.html}};
  \bibinfo{year}{2018}.
\newblock \bibinfo{note}{Accessed: 25/04/2018}.
\bibitem[{Keller(1997)}]{keller1997instant}
\bibinfo{author}{Keller\xfnm[ A]}.
\newblock \bibinfo{title}{Instant radiosity}.
\newblock In: \bibinfo{booktitle}{Proceedings of the 24th Annual Conference on
  Computer Graphics and Interactive Techniques}. SIGGRAPH '97;
  \bibinfo{address}{New York, NY, USA}: \bibinfo{publisher}{ACM
  Press/Addison-Wesley Publishing Co.}
\newblock ISBN \bibinfo{isbn}{0-89791-896-7}; \bibinfo{year}{1997}, p.
  \bibinfo{pages}{49--56}.
\newblock \URLprefix \url{https://doi.org/10.1145/258734.258769}.
  \DOIprefix\doi{10.1145/258734.258769}.
\bibitem[{Krivanek et~al.(2005)Krivanek, Gautron, Pattanaik and
  Bouatouch}]{krivanek2005radiance}
\bibinfo{author}{Krivanek\xfnm[ J]}, \bibinfo{author}{Gautron\xfnm[ P]},
  \bibinfo{author}{Pattanaik\xfnm[ S]}, \bibinfo{author}{Bouatouch\xfnm[ K]}.
\newblock \bibinfo{title}{Radiance caching for efficient global illumination
  computation}.
\newblock \bibinfo{journal}{IEEE Transactions on Visualization and Computer
  Graphics}
  \bibinfo{year}{2005};\bibinfo{volume}{11}(\bibinfo{number}{5}):\bibinfo{pages}{550--561}.
\bibitem[{Thomas and Forbes(2017)}]{thomas_deep_illumination}
\bibinfo{author}{Thomas\xfnm[ MM]}, \bibinfo{author}{Forbes\xfnm[ AG]}.
\newblock \bibinfo{title}{Deep illumination: Approximating dynamic global
  illumination with generative adversarial network}.
\newblock In: \bibinfo{booktitle}{arXiv preprint arXiv:1710.09834}.
  \bibinfo{year}{2017},.
\bibitem[{Chambolle and Pock(2011)}]{chambolle2011first}
\bibinfo{author}{Chambolle\xfnm[ A]}, \bibinfo{author}{Pock\xfnm[ T]}.
\newblock \bibinfo{title}{A first-order primal-dual algorithm for convex
  problems with applications to imaging}.
\newblock \bibinfo{journal}{Journal of mathematical imaging and vision}
  \bibinfo{year}{2011};\bibinfo{volume}{40}(\bibinfo{number}{1}):\bibinfo{pages}{120--145}.
\bibitem[{Beck and Teboulle(2009)}]{beck2009fast}
\bibinfo{author}{Beck\xfnm[ A]}, \bibinfo{author}{Teboulle\xfnm[ M]}.
\newblock \bibinfo{title}{Fast gradient-based algorithms for constrained total
  variation image denoising and deblurring problems}.
\newblock \bibinfo{journal}{IEEE transactions on image processing}
  \bibinfo{year}{2009};\bibinfo{volume}{18}(\bibinfo{number}{11}):\bibinfo{pages}{2419--2434}.
\bibitem[{Buades et~al.(2005)Buades, Coll and Morel}]{buades2005non}
\bibinfo{author}{Buades\xfnm[ A]}, \bibinfo{author}{Coll\xfnm[ B]},
  \bibinfo{author}{Morel\xfnm[ JM]}.
\newblock \bibinfo{title}{A non-local algorithm for image denoising}.
\newblock In: \bibinfo{booktitle}{2005 IEEE Computer Society Conference on
  Computer Vision and Pattern Recognition (CVPR'05)}; vol.~\bibinfo{volume}{2}.
  \bibinfo{organization}{IEEE}; \bibinfo{year}{2005}, p.
  \bibinfo{pages}{60--65}.
\bibitem[{Elad and Aharon(2006)}]{elad2006image}
\bibinfo{author}{Elad\xfnm[ M]}, \bibinfo{author}{Aharon\xfnm[ M]}.
\newblock \bibinfo{title}{Image denoising via sparse and redundant
  representations over learned dictionaries}.
\newblock \bibinfo{journal}{IEEE Transactions on Image processing}
  \bibinfo{year}{2006};\bibinfo{volume}{15}(\bibinfo{number}{12}):\bibinfo{pages}{3736--3745}.
\bibitem[{Liu et~al.(2018)Liu, Zhang, Zhang, Lin and Zuo}]{liu2018multi}
\bibinfo{author}{Liu\xfnm[ P]}, \bibinfo{author}{Zhang\xfnm[ H]},
  \bibinfo{author}{Zhang\xfnm[ K]}, \bibinfo{author}{Lin\xfnm[ L]},
  \bibinfo{author}{Zuo\xfnm[ W]}.
\newblock \bibinfo{title}{Multi-level wavelet-cnn for image restoration}.
\newblock In: \bibinfo{booktitle}{Proceedings of the IEEE Conference on
  Computer Vision and Pattern Recognition Workshops}. \bibinfo{year}{2018}, p.
  \bibinfo{pages}{773--782}.
\bibitem[{Bitterli et~al.(2016)Bitterli, Rousselle, Moon, Iglesias-Guiti{\'a}n,
  Adler, Mitchell et~al.}]{bitterli2016nonlinearly}
\bibinfo{author}{Bitterli\xfnm[ B]}, \bibinfo{author}{Rousselle\xfnm[ F]},
  \bibinfo{author}{Moon\xfnm[ B]}, \bibinfo{author}{Iglesias-Guiti{\'a}n\xfnm[
  JA]}, \bibinfo{author}{Adler\xfnm[ D]}, \bibinfo{author}{Mitchell\xfnm[ K]},
  et~al.
\newblock \bibinfo{title}{Nonlinearly weighted first-order regression for
  denoising monte carlo renderings}.
\newblock In: \bibinfo{booktitle}{Computer Graphics Forum};
  vol.~\bibinfo{volume}{35}. \bibinfo{organization}{Wiley Online Library};
  \bibinfo{year}{2016}, p. \bibinfo{pages}{107--117}.
\bibitem[{Moon et~al.(2016)Moon, McDonagh, Mitchell and
  Gross}]{moon2016adaptive}
\bibinfo{author}{Moon\xfnm[ B]}, \bibinfo{author}{McDonagh\xfnm[ S]},
  \bibinfo{author}{Mitchell\xfnm[ K]}, \bibinfo{author}{Gross\xfnm[ M]}.
\newblock \bibinfo{title}{Adaptive polynomial rendering}.
\newblock \bibinfo{journal}{ACM Transactions on Graphics (TOG)}
  \bibinfo{year}{2016};\bibinfo{volume}{35}(\bibinfo{number}{4}):\bibinfo{pages}{1--10}.
\bibitem[{Kalantari et~al.(2015)Kalantari, Bako and
  Sen}]{kalantari_machine_learning_ml_noise}
\bibinfo{author}{Kalantari\xfnm[ NK]}, \bibinfo{author}{Bako\xfnm[ S]},
  \bibinfo{author}{Sen\xfnm[ P]}.
\newblock \bibinfo{title}{A machine learning approach for filtering monte carlo
  noise}.
\newblock \bibinfo{journal}{ACM Trans Graph}
  \bibinfo{year}{2015};\bibinfo{volume}{34}(\bibinfo{number}{4}):\bibinfo{pages}{122:1--122:12}.
\newblock \URLprefix \url{http://doi.acm.org/10.1145/2766977}.
  \DOIprefix\doi{10.1145/2766977}.
\bibitem[{Pharr and Humphreys(2010)}]{pbrt2}
\bibinfo{author}{Pharr\xfnm[ M]}, \bibinfo{author}{Humphreys\xfnm[ G]}.
\newblock \bibinfo{title}{Physically Based Rendering: From Theory To
  Implementation}.
\newblock \bibinfo{publisher}{Morgan Kaufmann, 2nd edition};
  \bibinfo{year}{2010}.
\bibitem[{Bako et~al.(2017)Bako, Vogels, McWilliams, Meyer, Novák, Harvill
  et~al.}]{kernel_predicting_denoising_2017}
\bibinfo{author}{Bako\xfnm[ S]}, \bibinfo{author}{Vogels\xfnm[ T]},
  \bibinfo{author}{McWilliams\xfnm[ B]}, \bibinfo{author}{Meyer\xfnm[ M]},
  \bibinfo{author}{Novák\xfnm[ J]}, \bibinfo{author}{Harvill\xfnm[ A]}, et~al.
\newblock \bibinfo{title}{Kernel-predicting convolutional networks for
  denoising monte carlo renderings}.
\newblock In: \bibinfo{booktitle}{Proceedings of ACM SIGGRAPH 2017}.
  \bibinfo{publisher}{ACM}; \bibinfo{year}{2017},.
\bibitem[{Vogels et~al.(2018)Vogels, Rousselle, McWilliams, R\"othlin, Harvill,
  Adler et~al.}]{Vogels2018KPAL}
\bibinfo{author}{Vogels\xfnm[ T]}, \bibinfo{author}{Rousselle\xfnm[ F]},
  \bibinfo{author}{McWilliams\xfnm[ B]}, \bibinfo{author}{R\"othlin\xfnm[ G]},
  \bibinfo{author}{Harvill\xfnm[ A]}, \bibinfo{author}{Adler\xfnm[ D]}, et~al.
\newblock \bibinfo{title}{Denoising with kernel prediction and asymmetric loss
  functions}.
\newblock \bibinfo{journal}{ACM Transactions on Graphics (Proceedings of
  SIGGRAPH 2018)}
  \bibinfo{year}{2018};\bibinfo{volume}{37}(\bibinfo{number}{4}):\bibinfo{pages}{124:1--124:15}.
\newblock \DOIprefix\doi{10.1145/3197517.3201388}.
\bibitem[{Chaitanya et~al.(2017)Chaitanya, Kaplanyan, Schied, Salvi, Lefohn,
  Nowrouzezahrai et~al.}]{chaitanya2017_interactive_reconstruction}
\bibinfo{author}{Chaitanya\xfnm[ CRA]}, \bibinfo{author}{Kaplanyan\xfnm[ AS]},
  \bibinfo{author}{Schied\xfnm[ C]}, \bibinfo{author}{Salvi\xfnm[ M]},
  \bibinfo{author}{Lefohn\xfnm[ A]}, \bibinfo{author}{Nowrouzezahrai\xfnm[ D]},
  et~al.
\newblock \bibinfo{title}{Interactive reconstruction of monte carlo image
  sequences using a recurrent denoising autoencoder}.
\newblock \bibinfo{journal}{ACM Trans Graph}
  \bibinfo{year}{2017};\bibinfo{volume}{36}(\bibinfo{number}{4}):\bibinfo{pages}{98:1--98:12}.
\newblock \URLprefix \url{http://doi.acm.org/10.1145/3072959.3073601}.
  \DOIprefix\doi{10.1145/3072959.3073601}.
\bibitem[{Kuznetsov et~al.(2018)Kuznetsov, Kalantari and
  Ramamoorthi}]{kuznetsov2018deep}
\bibinfo{author}{Kuznetsov\xfnm[ A]}, \bibinfo{author}{Kalantari\xfnm[ NK]},
  \bibinfo{author}{Ramamoorthi\xfnm[ R]}.
\newblock \bibinfo{title}{Deep adaptive sampling for low sample count
  rendering}.
\newblock In: \bibinfo{booktitle}{Computer Graphics Forum};
  vol.~\bibinfo{volume}{37}. \bibinfo{organization}{Wiley Online Library};
  \bibinfo{year}{2018}, p. \bibinfo{pages}{35--44}.
\bibitem[{Vicini et~al.(2019)Vicini, Adler, Nov{\'a}k, Rousselle and
  Burley}]{vicini2019denoising}
\bibinfo{author}{Vicini\xfnm[ D]}, \bibinfo{author}{Adler\xfnm[ D]},
  \bibinfo{author}{Nov{\'a}k\xfnm[ J]}, \bibinfo{author}{Rousselle\xfnm[ F]},
  \bibinfo{author}{Burley\xfnm[ B]}.
\newblock \bibinfo{title}{Denoising deep monte carlo renderings}.
\newblock In: \bibinfo{booktitle}{Computer Graphics Forum};
  vol.~\bibinfo{volume}{38}. \bibinfo{organization}{Wiley Online Library};
  \bibinfo{year}{2019}, p. \bibinfo{pages}{316--327}.
\bibitem[{Gharbi et~al.(2019)Gharbi, Li, Aittala, Lehtinen and
  Durand}]{gharbi2019sample}
\bibinfo{author}{Gharbi\xfnm[ M]}, \bibinfo{author}{Li\xfnm[ TM]},
  \bibinfo{author}{Aittala\xfnm[ M]}, \bibinfo{author}{Lehtinen\xfnm[ J]},
  \bibinfo{author}{Durand\xfnm[ F]}.
\newblock \bibinfo{title}{Sample-based monte carlo denoising using a
  kernel-splatting network}.
\newblock \bibinfo{journal}{ACM Transactions on Graphics (TOG)}
  \bibinfo{year}{2019};\bibinfo{volume}{38}(\bibinfo{number}{4}):\bibinfo{pages}{1--12}.
\bibitem[{Xu et~al.(2019)Xu, Zhang, Wang, Xu, Yang, Li
  et~al.}]{xu2019adversarial}
\bibinfo{author}{Xu\xfnm[ B]}, \bibinfo{author}{Zhang\xfnm[ J]},
  \bibinfo{author}{Wang\xfnm[ R]}, \bibinfo{author}{Xu\xfnm[ K]},
  \bibinfo{author}{Yang\xfnm[ YL]}, \bibinfo{author}{Li\xfnm[ C]}, et~al.
\newblock \bibinfo{title}{Adversarial monte carlo denoising with conditioned
  auxiliary feature modulation}.
\newblock \bibinfo{journal}{ACM Transactions on Graphics (TOG)}
  \bibinfo{year}{2019};\bibinfo{volume}{38}(\bibinfo{number}{6}):\bibinfo{pages}{1--12}.
\bibitem[{Ledig et~al.(2017)Ledig, Theis, Husz{\'a}r, Caballero, Cunningham,
  Acosta et~al.}]{ledig2017photo}
\bibinfo{author}{Ledig\xfnm[ C]}, \bibinfo{author}{Theis\xfnm[ L]},
  \bibinfo{author}{Husz{\'a}r\xfnm[ F]}, \bibinfo{author}{Caballero\xfnm[ J]},
  \bibinfo{author}{Cunningham\xfnm[ A]}, \bibinfo{author}{Acosta\xfnm[ A]},
  et~al.
\newblock \bibinfo{title}{Photo-realistic single image super-resolution using a
  generative adversarial network}.
\newblock In: \bibinfo{booktitle}{Proceedings of the IEEE conference on
  computer vision and pattern recognition}. \bibinfo{year}{2017}, p.
  \bibinfo{pages}{4681--4690}.
\bibitem[{Nalbach et~al.(2017)Nalbach, Arabadzhiyska, Mehta, Seidel and
  Ritschel}]{nalbach_deep_shading}
\bibinfo{author}{Nalbach\xfnm[ O]}, \bibinfo{author}{Arabadzhiyska\xfnm[ E]},
  \bibinfo{author}{Mehta\xfnm[ D]}, \bibinfo{author}{Seidel\xfnm[ HP]},
  \bibinfo{author}{Ritschel\xfnm[ T]}.
\newblock \bibinfo{title}{Deep shading: Convolutional neural networks for
  screen-space shading}.
\newblock In: \bibinfo{booktitle}{Computer Graphics Forum}.
  \bibinfo{year}{2017},.
\bibitem[{Ren et~al.(2013)Ren, Wang, Gong, Lin, Tong and Guo}]{ren2013global}
\bibinfo{author}{Ren\xfnm[ P]}, \bibinfo{author}{Wang\xfnm[ J]},
  \bibinfo{author}{Gong\xfnm[ M]}, \bibinfo{author}{Lin\xfnm[ S]},
  \bibinfo{author}{Tong\xfnm[ X]}, \bibinfo{author}{Guo\xfnm[ B]}.
\newblock \bibinfo{title}{Global illumination with radiance regression
  functions}.
\newblock \bibinfo{journal}{ACM Transactions on Graphics (TOG)}
  \bibinfo{year}{2013};\bibinfo{volume}{32}(\bibinfo{number}{4}):\bibinfo{pages}{130}.
\bibitem[{Dahm and Keller(2016)}]{dahm2016learning}
\bibinfo{author}{Dahm\xfnm[ K]}, \bibinfo{author}{Keller\xfnm[ A]}.
\newblock \bibinfo{title}{Learning light transport the reinforced way}.
\newblock In: \bibinfo{booktitle}{International Conference on Monte Carlo and
  Quasi-Monte Carlo Methods in Scientific Computing}.
  \bibinfo{organization}{Springer Proceedings in Mathematics and Statistics Vol
  241}; \bibinfo{year}{2016},.
\bibitem[{Dahm and Keller(2017)}]{integralEquations}
\bibinfo{author}{Dahm\xfnm[ K]}, \bibinfo{author}{Keller\xfnm[ A]}.
\newblock \bibinfo{title}{Machine learning and integral equations}.
\newblock \bibinfo{journal}{CoRR}
  \bibinfo{year}{2017};\bibinfo{volume}{abs/1712.06115}.
\newblock \URLprefix \url{http://arxiv.org/abs/1712.06115}.
  \href{http://arxiv.org/abs/1712.06115}{\tt arXiv:1712.06115}.
\bibitem[{Vorba et~al.(2014)Vorba, Karl{\'i}k, {\v{S}}ik, Ritschel and
  K{\v{r}}iv{\'{a}}nek}]{parametricMixtureLt}
\bibinfo{author}{Vorba\xfnm[ J]}, \bibinfo{author}{Karl{\'i}k\xfnm[ O]},
  \bibinfo{author}{{\v{S}}ik\xfnm[ M]}, \bibinfo{author}{Ritschel\xfnm[ T]},
  \bibinfo{author}{K{\v{r}}iv{\'{a}}nek\xfnm[ J]}.
\newblock \bibinfo{title}{On-line learning of parametric mixture models for
  light transport simulation}.
\newblock \bibinfo{journal}{ACM Transactions on Graphics (Proceedings of
  SIGGRAPH 2014)}
  \bibinfo{year}{2014};\bibinfo{volume}{33}(\bibinfo{number}{4}).
\bibitem[{Herholz et~al.(2016)Herholz, Elek, Vorba, Lensch and
  K{\v{r}}iv{\'a}nek}]{herholz2016product}
\bibinfo{author}{Herholz\xfnm[ S]}, \bibinfo{author}{Elek\xfnm[ O]},
  \bibinfo{author}{Vorba\xfnm[ J]}, \bibinfo{author}{Lensch\xfnm[ H]},
  \bibinfo{author}{K{\v{r}}iv{\'a}nek\xfnm[ J]}.
\newblock \bibinfo{title}{Product importance sampling for light transport path
  guiding}.
\newblock In: \bibinfo{booktitle}{Computer Graphics Forum};
  vol.~\bibinfo{volume}{35}. \bibinfo{organization}{Wiley Online Library};
  \bibinfo{year}{2016}, p. \bibinfo{pages}{67--77}.
\bibitem[{M{\"u}ller et~al.(2017)M{\"u}ller, Gross and
  Nov{\'a}k}]{muller2017practical}
\bibinfo{author}{M{\"u}ller\xfnm[ T]}, \bibinfo{author}{Gross\xfnm[ M]},
  \bibinfo{author}{Nov{\'a}k\xfnm[ J]}.
\newblock \bibinfo{title}{Practical path guiding for efficient light-transport
  simulation}.
\newblock In: \bibinfo{booktitle}{Computer Graphics Forum};
  vol.~\bibinfo{volume}{36}. \bibinfo{organization}{Wiley Online Library};
  \bibinfo{year}{2017}, p. \bibinfo{pages}{91--100}.
\bibitem[{M\"{u}ller(2019)}]{mueller19guiding}
\bibinfo{author}{M\"{u}ller\xfnm[ T]}.
\newblock \bibinfo{title}{``practical path guiding'' in production}.
\newblock In: \bibinfo{booktitle}{ACM SIGGRAPH Courses: Path Guiding in
  Production, Chapter 10}. \bibinfo{address}{New York, NY, USA}:
  \bibinfo{publisher}{ACM}; \bibinfo{year}{2019}, p.
  \bibinfo{pages}{18:35--18:48}.
\newblock \DOIprefix\doi{10.1145/3305366.3328091}.
\bibitem[{Kallweit et~al.(2017)Kallweit, M{\"u}ller, McWilliams, Gross and
  Nov{\'a}k}]{kallweit2017deep}
\bibinfo{author}{Kallweit\xfnm[ S]}, \bibinfo{author}{M{\"u}ller\xfnm[ T]},
  \bibinfo{author}{McWilliams\xfnm[ B]}, \bibinfo{author}{Gross\xfnm[ M]},
  \bibinfo{author}{Nov{\'a}k\xfnm[ J]}.
\newblock \bibinfo{title}{Deep scattering: Rendering atmospheric clouds with
  radiance-predicting neural networks}.
\newblock \bibinfo{journal}{ACM Transactions on Graphics (TOG)}
  \bibinfo{year}{2017};\bibinfo{volume}{36}(\bibinfo{number}{6}):\bibinfo{pages}{231}.
\bibitem[{Kajiya(1998)}]{kajiya1986rendering}
\bibinfo{author}{Kajiya\xfnm[ JT]}.
\newblock \bibinfo{title}{Seminal graphics}.
\newblock chap. \bibinfo{chapter}{The Rendering Equation}.
  \bibinfo{address}{New York, NY, USA}: \bibinfo{publisher}{ACM}.
\newblock ISBN \bibinfo{isbn}{1-58113-052-X}; \bibinfo{year}{1998}, p.
  \bibinfo{pages}{157--164}.
\newblock \URLprefix \url{http://doi.acm.org/10.1145/280811.280987}.
  \DOIprefix\doi{10.1145/280811.280987}.
\bibitem[{Veach and Guibas(1995)}]{veach1995bidirectional_estimators}
\bibinfo{author}{Veach\xfnm[ E]}, \bibinfo{author}{Guibas\xfnm[ L]}.
\newblock \bibinfo{title}{Bidirectional estimators for light transport}.
\newblock In: \bibinfo{booktitle}{Photorealistic Rendering Techniques}.
  \bibinfo{publisher}{Springer}; \bibinfo{year}{1995}, p.
  \bibinfo{pages}{145--167}.
\bibitem[{Ronneberger et~al.(2015)Ronneberger, Fischer and
  Brox}]{ronneberger2015u}
\bibinfo{author}{Ronneberger\xfnm[ O]}, \bibinfo{author}{Fischer\xfnm[ P]},
  \bibinfo{author}{Brox\xfnm[ T]}.
\newblock \bibinfo{title}{{U-Net}: Convolutional networks for biomedical image
  segmentation}.
\newblock In: \bibinfo{booktitle}{International Conference on Medical image
  computing and computer-assisted intervention}.
  \bibinfo{organization}{Springer}; \bibinfo{year}{2015}, p.
  \bibinfo{pages}{234--241}.
\bibitem[{Bitterli(2018)}]{benedikt_scenes}
\bibinfo{author}{Bitterli\xfnm[ B]}.
\newblock \bibinfo{title}{Rendering resources}.
\newblock
  \bibinfo{howpublished}{\url{https://benedikt-bitterli.me/resources/}};
  \bibinfo{year}{2018}.
\newblock \bibinfo{note}{Accessed: 24/05/2018}.
\bibitem[{Pharr et~al.(2016)Pharr, Jakob and Humphreys}]{pbrt}
\bibinfo{author}{Pharr\xfnm[ M]}, \bibinfo{author}{Jakob\xfnm[ W]},
  \bibinfo{author}{Humphreys\xfnm[ G]}.
\newblock \bibinfo{title}{Physically Based Rendering: From Theory To
  Implementation}.
\newblock \bibinfo{publisher}{Morgan Kaufmann, 3rd edition};
  \bibinfo{year}{2016}.
\bibitem[{Whittle et~al.(2017)Whittle, Jones and Mantiuk}]{whittle2017analysis}
\bibinfo{author}{Whittle\xfnm[ J]}, \bibinfo{author}{Jones\xfnm[ MW]},
  \bibinfo{author}{Mantiuk\xfnm[ R]}.
\newblock \bibinfo{title}{Analysis of reported error in monte carlo rendered
  images}.
\newblock \bibinfo{journal}{The Visual Computer}
  \bibinfo{year}{2017};\bibinfo{volume}{33}(\bibinfo{number}{6-8}):\bibinfo{pages}{705--713}.
\bibitem[{Lehtinen et~al.(2018)Lehtinen, Munkberg, Hasselgren, Laine, Karras,
  Aittala et~al.}]{lehtinen2018noise2noise}
\bibinfo{author}{Lehtinen\xfnm[ J]}, \bibinfo{author}{Munkberg\xfnm[ J]},
  \bibinfo{author}{Hasselgren\xfnm[ J]}, \bibinfo{author}{Laine\xfnm[ S]},
  \bibinfo{author}{Karras\xfnm[ T]}, \bibinfo{author}{Aittala\xfnm[ M]}, et~al.
\newblock \bibinfo{title}{Noise2noise: Learning image restoration without clean
  data}.
\newblock \bibinfo{journal}{arXiv180304189} \bibinfo{year}{2018};.
\bibitem[{Wang et~al.(2004)Wang, Bovik, Sheikh and Simoncelli}]{wangSsim2004}
\bibinfo{author}{Wang\xfnm[ Z]}, \bibinfo{author}{Bovik\xfnm[ AC]},
  \bibinfo{author}{Sheikh\xfnm[ HR]}, \bibinfo{author}{Simoncelli\xfnm[ EP]}.
\newblock \bibinfo{title}{Image quality assessment: from error visibility to
  structural similarity}.
\newblock \bibinfo{journal}{IEEE Transactions on Image Processing}
  \bibinfo{year}{2004};\bibinfo{volume}{13}(\bibinfo{number}{4}):\bibinfo{pages}{600--612}.
\bibitem[{Kruskal and Wallis(1952)}]{kruskal1952use}
\bibinfo{author}{Kruskal\xfnm[ WH]}, \bibinfo{author}{Wallis\xfnm[ WA]}.
\newblock \bibinfo{title}{Use of ranks in one-criterion variance analysis}.
\newblock \bibinfo{journal}{Journal of the American statistical Association}
  \bibinfo{year}{1952};\bibinfo{volume}{47}(\bibinfo{number}{260}):\bibinfo{pages}{583--621}.
\bibitem[{Cenobi(2012)}]{cenobi_bathroom_green}
\bibinfo{author}{Cenobi\xfnm[ U]}.
\newblock \bibinfo{title}{Bathroom scene}.
\newblock
  \bibinfo{howpublished}{\url{https://www.blendswap.com/blends/view/52486}};
  \bibinfo{year}{2012}.
\newblock \bibinfo{note}{Accessed: 29/08/2018}.
\bibitem[{Intel\textsuperscript{\textregistered}(2020)}]{Intel2020}
\bibinfo{author}{Intel\textsuperscript{\textregistered}\xfnm[]}.
\newblock \bibinfo{title}{Open image denoiser}.
\newblock \bibinfo{type}{Tech. Rep.}; Intel (R); \bibinfo{year}{2020}.
\newblock \URLprefix \url{https://openimagedenoise.github.io/}.
\bibitem[{Parker et~al.(2010)Parker, Bigler, Dietrich, Friedrich, Hoberock,
  Luebke et~al.}]{parker2010optix}
\bibinfo{author}{Parker\xfnm[ SG]}, \bibinfo{author}{Bigler\xfnm[ J]},
  \bibinfo{author}{Dietrich\xfnm[ A]}, \bibinfo{author}{Friedrich\xfnm[ H]},
  \bibinfo{author}{Hoberock\xfnm[ J]}, \bibinfo{author}{Luebke\xfnm[ D]},
  et~al.
\newblock \bibinfo{title}{Optix: a general purpose ray tracing engine}.
\newblock \bibinfo{journal}{ACM Transactions on Graphics (TOG)}
  \bibinfo{year}{2010};\bibinfo{volume}{29}(\bibinfo{number}{4}):\bibinfo{pages}{66}.
\bibitem[{Goodfellow et~al.(2016)Goodfellow, Bengio, Courville and
  Bengio}]{goodfellow2016deep}
\bibinfo{author}{Goodfellow\xfnm[ I]}, \bibinfo{author}{Bengio\xfnm[ Y]},
  \bibinfo{author}{Courville\xfnm[ A]}, \bibinfo{author}{Bengio\xfnm[ Y]}.
\newblock \bibinfo{title}{Deep learning}; vol.~\bibinfo{volume}{1}.
\newblock \bibinfo{publisher}{MIT press Cambridge}; \bibinfo{year}{2016}.
\bibitem[{Mara et~al.(2017)Mara, McGuire, Bitterli and
  Jarosz}]{mara2017efficient}
\bibinfo{author}{Mara\xfnm[ M]}, \bibinfo{author}{McGuire\xfnm[ M]},
  \bibinfo{author}{Bitterli\xfnm[ B]}, \bibinfo{author}{Jarosz\xfnm[ W]}.
\newblock \bibinfo{title}{An efficient denoising algorithm for global
  illumination.}
\newblock \bibinfo{journal}{High Performance Graphics}
  \bibinfo{year}{2017};\bibinfo{volume}{10}:\bibinfo{pages}{3105762--3105774}.
\bibitem[{Burgess(2020)}]{burgess2020rtx}
\bibinfo{author}{Burgess\xfnm[ J]}.
\newblock \bibinfo{title}{{RTX on—The NVIDIA Turing GPU}}.
\newblock \bibinfo{journal}{IEEE Micro}
  \bibinfo{year}{2020};\bibinfo{volume}{40}(\bibinfo{number}{2}):\bibinfo{pages}{36--44}.

\end{thebibliography}

\end{document}